\documentclass[reprint,prb,longbibliography,superscriptaddress]{revtex4-2}

\usepackage{amssymb, amsmath}
\usepackage[english]{babel}
\usepackage{bm}
\usepackage{graphicx}
\usepackage[utf8]{inputenc}
\usepackage{bbold}

\usepackage[dvipsnames]{xcolor}

\newcommand{\commenttoggle}[1]{}

\newcommand{\red}[1]{{ #1}}

\newcommand{\apprefA}{Appendix \ref{app1}}
\newcommand{\apprefB}{Appendix \ref{app2}}
\newcommand{\apprefC}{Appendix \ref{app3}}

\mathchardef\myminus="2D

\newcommand\mydots{...\,}

\usepackage{letltxmacro}
\LetLtxMacro{\ORIGselectlanguage}{\selectlanguage}
\makeatletter
\DeclareRobustCommand{\selectlanguage}[1]{%
  \@ifundefined{alias@\string#1}
    {\ORIGselectlanguage{#1}}
    {\begingroup\edef\x{\endgroup
       \noexpand\ORIGselectlanguage{\@nameuse{alias@#1}}}\x}%
}
\newcommand{\definelanguagealias}[2]{%
  \@namedef{alias@#1}{#2}%
}

\makeatother

\definelanguagealias{en}{english}
\definelanguagealias{EN}{english}
\definelanguagealias{eng}{english}
\definelanguagealias{de}{ngerman}

\begin{document}
\author{Robert Fuchs}
\author{Marten Richter}
\email[]{marten.richter@tu-berlin.de}

\affiliation{Institut für Theoretische Physik, Nichtlineare Optik und
Quantenelektronik, Technische Universität Berlin, Hardenbergstr. 36, EW 7-1, 10623
Berlin, Germany}

\title{A hierarchical equations of motion (HEOM) analog for systems with delay: illustrated on inter-cavity photon propagation }



\begin{abstract}
Over the last two decades, the hierarchical equations of motion (HEOM) of Tanimura and Kubo have become the equation of motion-based tool for numerically exact calculations of system-bath problems. The HEOM is today generalized to many cases of dissipation and transfer processes through an external bath. In spatially extended photonic systems, the propagation of photons through the bath leads to retardation/delays in the coupling of quantum emitters. Here, the idea behind the HEOM derivation is generalized to the case of photon retardation and applied to the simple example of two dielectric slabs. The derived equations provide a simple reliable framework for describing retardation and may provide an alternative to path integral treatments.
\end{abstract}


\date{\today}
\maketitle
\section{Introduction}
After the hierarchical equations of motion (HEOM) were initially invented by Tanimura and Kubo \cite{originalheom,tanimura2020numerically} to solve numerically exactly the open quantum system problem with a Debye spectral density, the HEOM did not immediately take off, since the limited numeric capabilities did not allow for a versatile implementation at the time. However, the idea of using the time constant derivative of the Debye spectral density time correlation function stuck.
Recently, various implementations \cite{heomquick,lambert2020bofin,kramer2018efficient,seibt2021strong,tanimura2020numerically,kramer2018energy} of HEOM followed after sufficient computing power became available. Soon after its invention, many generalizations using arbitrary spectral densities by decomposition into summed Debye form spectral densities were also developed. For most system-bath approaches it provides a well-established path to a numerically exact solution.

A different type of system-bath problem is the propagation of quantum states, e.g., through a bath of photons or phonons \cite{oulton2008hybrid,stockman2004nanofocusing,orieux2017semiconductor,weiss2018interfacing,jayakumar2014time,CarmeleReitzenstein,PhysRevLett.116.093601,kaestle2020protected,PhysRevResearch.3.023030, PhysRevLett.128.167403}. A typical problem is describing quantum interconnects for quantum computing and cryptography applications. 
Recently, various applications of these systems with a delay caused by the propagation through the bath were investigated \cite{oulton2008hybrid,stockman2004nanofocusing,orieux2017semiconductor,weiss2018interfacing,jayakumar2014time,CarmeleReitzenstein,PhysRevLett.116.093601,kaestle2020protected,PhysRevResearch.3.023030, PhysRevLett.128.167403} including the development of different methods.
However, the number of propagating photons is still limited, as it was for the open quantum systems approaches until HEOM implementations became widespread, along with other methods such as tensor networks \cite{CALDEIRA1983587,tanimura1993real,makri1995tensor,makri1995tensor2,vagov2011real,strathearn2017efficient,strathearn2018TEMPO,gribben2021exact,cygorek2021numerically,PhysRevLett.116.093601,kaestle2020protected,PhysRevA.87.013428,caycedosoler2021exact,plenioheisenberg,PhysRevLett.116.237201,Rosenbach_2016,PhysRevLett.116.093601,kaestle2020protected,schroder2019tensor,PhysRevLett.123.100502}.
In this paper, an analysis of the HEOM derivation in the context of delay is carried out and HEOM analog equations for systems with delay are derived.
We demonstrate that the approach leads to a systematic set of equations ordered by the number of photons propagating through the bath. In the future, combinations with, e.g., tensor networks or automatic derivation may lead to an additional route to solve problems involving delays.

The paper starts with a derivation of the HEOM analog for open quantum systems with delay and illustrates its potential with a simple photon propagation example.

\section{Derivation of hierachical equations of motion (HEOM)}
An HEOM analog with delay is derived for an open quantum system with: $H=H_{s}+H_b+H_{sb}$.
Here, $H_s$ is the Hamiltonian of the system, which consists of quantum emitters in different spatially separated cavities. $H_b$ is the bath Hamiltonian containing the propagating photon modes. Finally, $H_{sb}$ is the system-bath coupling Hamiltonian.
In open quantum systems, only the observables of the system are of interest, which can be calculated from the relevant density matrix $\rho_s(t)=\mathrm{tr}_B(\rho(t))$. Its calculation is the main objective of HEOM, where we transfer the steps by Tanimura and Kubo \cite{originalheom} to systems with delay.
We assume a factorized initial state $\rho(t_0)=\rho_s(t_0)\otimes \rho_B$, where $\rho_B$ is a harmonic bath state.
The system dynamics obey:
\begin{align}
&\rho_s(t) = \mathrm{tr}_B \left(T_\leftarrow U(t,t_0)\right. \nonumber\\
&\qquad\mathrm{exp}\left( -\frac{i}{\hbar} \int_{t_0}^t \mathrm{d} \tau U(t_0,\tau) H_{sb,-}(\tau)U(\tau, t_0)\right)\nonumber\\ &\qquad\quad\left.\rho_s(t_0)\otimes \rho_B\right), \label{sys_density}
\end{align}
where  $A_L\rho=A\rho$, $A_R\rho=\rho A$, and $A_-=A_L-A_R$ define the Liouville space operators acting on Liouville operator $\rho$ for any Hilbert space operator $A$ \cite{Chernyak:1996} and $U(t,t_0)=T_\leftarrow \mathrm{exp}\left( -\frac{i}{\hbar} \int_{t_0}^t \mathrm{d} \tau  (H_{s,-}(\tau)+H_{b,-}(\tau))\right)$ with time ordering operator $T_\leftarrow$. $H_{s,-}(\tau)$ may also contain Lindblad operators for describing external processes acting on the joint system-bath state.
Following the HEOM derivation \cite{originalheom}  and the path integral derivation from \cite{PhysRevLett.128.167403}, we convert Eq.~\eqref{sys_density} to path integral form:
\begin{align}
&\rho_s(t)=\mathrm{tr}_B\left( \right.\nonumber\\
& T_\leftarrow\prod_{i=0}^M U_{i,i-1} \mathrm{exp}\left(\int_{t_0+\varepsilon(i-1)}^{t_0+\varepsilon i}\mathrm{d}\tau U^\dagger_{i-1}(\tau)H_{sb,-}(\tau)U_{i-1}(\tau)\right)\nonumber \\ &\qquad\quad\left.\rho_s(t_0)\otimes\rho_B\right)
\end{align}
with $\varepsilon=(t-t_0)/M$ and $M\rightarrow\infty$ (in the following equations the limit is always assumed). Furthermore, $U_{i,j}=U(t_0+\varepsilon i, t_0+\varepsilon j)$ and $U_{i}(\tau)= U(\tau, t_0+\varepsilon(i))$.
For small $\varepsilon$, the approximation $U_{i,i-1}~\mathrm{exp}\left(\int_{t_0+\varepsilon(i-1)}^{t_0+\varepsilon i}\mathrm{d}\tau U_{i-1}^\dagger(\tau)H_{sb}(\tau)U_{i-1}(\tau)\right) \approx U_{i,i-1} + \varepsilon \cdot U_{i,i-1/2} H_{sb}(t_0+\varepsilon(i-1/2)) U_{i-1/2,i-1} =: U_{i,i -1} + \varepsilon \cdot U_{sb}^{(1)}(i)$ holds, yielding:
\begin{align}
\rho_s(t)=\mathrm{tr}_B\left(T_\leftarrow \prod_{i=0}^M (U_{i,i-1} + \varepsilon \cdot U_{sb}^{(1)}(i)) \rho_s(t_0)\otimes\rho_B\right). \nonumber
\end{align}
\begin{figure}[b]
    \centering
    \includegraphics[width=8.6cm]{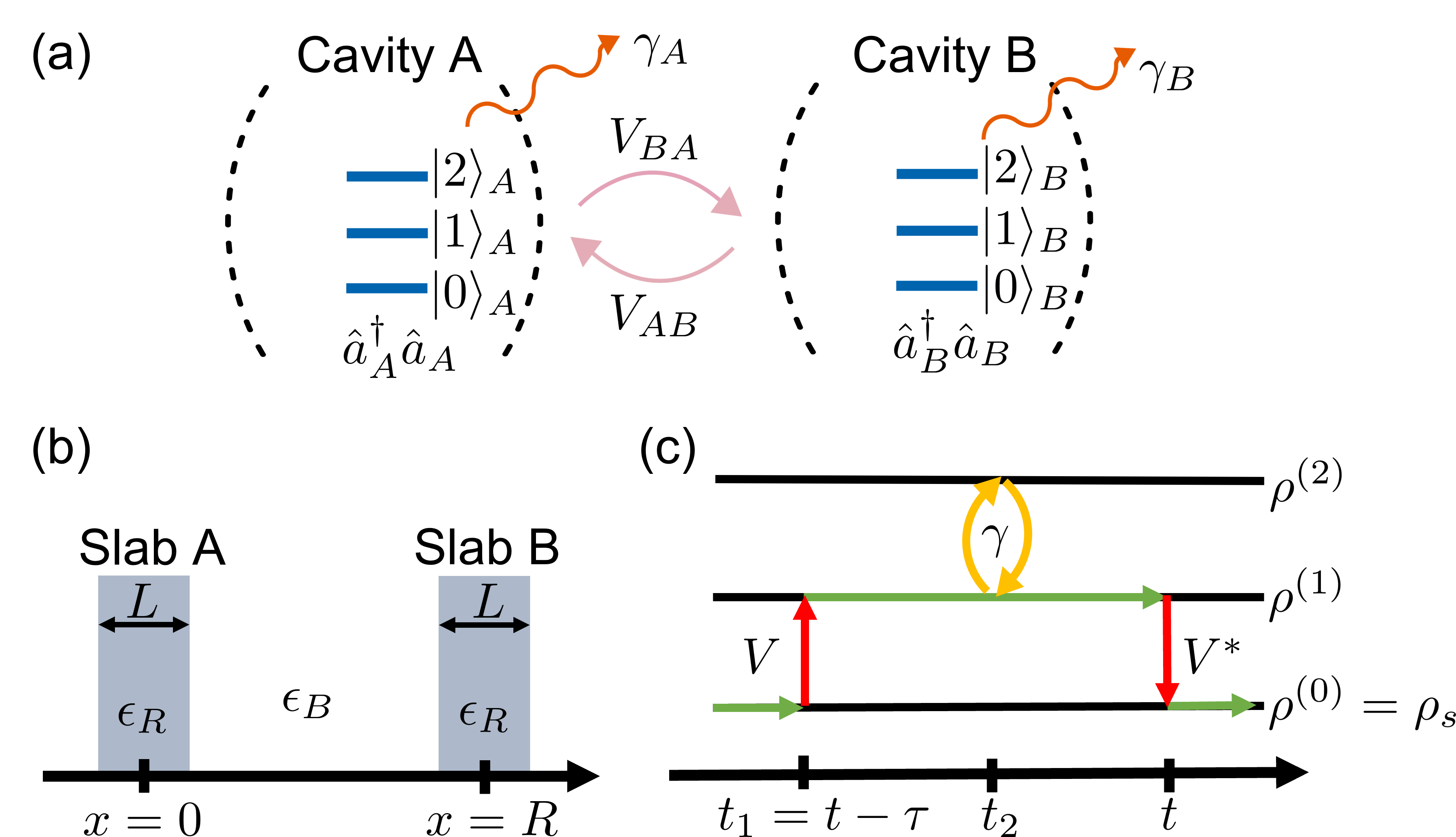}
    \caption{(a) Model of two open QNM cavities with dissipation rates \(\gamma_{\mu}\) and effective inter-cavity coupling strength \(V_{\mu\eta}\). (b) 1D model with two slabs with constant permittivity \(\epsilon_R = \pi^2\) serving as QNM cavities, sitting against a background \(\epsilon_B = 1\). (c) Scheme of the HEOM depicting a process including inter-cavity transfer and dissipation.}
    \label{fig:model}
\end{figure}
We assume linear system-bath coupling: $H_{sb}=\sum_{ij\mu} C_{ij\mu} A_{ij} B_{\mu}$ with system  $A_{ij}$ and linear bath operator $B_{\mu}$. For a system $A$ and bath $B$ Liouville operator the relation $(AB)_-=A_+B_-+A_-B_+$ holds, so $U_{sb}^{(1)}(i)$ can be written as a sum over products of the system and bath operators $U_{sb}^{(1)}(i)=\sum_l A_l^{(1)}(i) B_l^{(1)}(i)$, and we define $A_l^{(0)}=U^{s}_{i,i-1}\delta_{l,0}$ and $B_l^{(0)}=U^{b}_{i,i-1}\delta_{l,0}$ with the system and bath parts of $U_{i,i-1}$.
With these relations, we write $\rho_S$ in terms of a system part $S$ and an influence functional (similar form as in \cite{PhysRevLett.128.167403}),
\begin{align}
&\rho_s(t)= \sum_{k_1\mydots k_M=0}^1\sum_{ l_1\mydots l_M}  \left(\prod_{i=1}^M \varepsilon^{k_i}\right) S(k_1l_1,\mydots, k_M l_M) \nonumber\\
&\qquad\qquad \qquad \times I(k_1l_1,\mydots, k_M l_M).
\end{align}
The system part is still an operator $S(k_1l_1,\mydots, k_M l_M)=T_\leftarrow \prod_{i=1}^M A^{(k_i)}_{l_i}(i) \rho_s(t_0)$, while the influence functional $I(k_1l_1,\mydots, k_M l_M)=\mathrm{tr}_B(T_\leftarrow \prod_{i=1}^M B^{(k_i)}_{l_i}(i) \rho_B) $ is just a number.
Since $\rho_B$ is assumed to be a harmonic bath equilibrium state, Wick's theorem allows us to factorize the influence functional $I$ into expectation values of two bath operators $B_l^{(1)}(\cdot)$.
Furthermore, for small $\varepsilon$, the system propagator is roughly $U^{s}_{i,i-1}\approx Id_s -\frac{i}{\hbar} \varepsilon H_{s,-} (t_0+\varepsilon (i-1/2)) $. 
Using the approximations of the time propagators and using Wick's theorem we obtain, 
\begin{align}
&\rho_s(t+\varepsilon)\approx \rho_s(t)
-\varepsilon \frac{i}{\hbar} H_{s,-}(t_0+\varepsilon(M+1/2))\rho_s(t)\nonumber\\ &\quad+ \sum_{l_{M+1}} T_\leftarrow \varepsilon A_{l_{M+1}}^{(1)} \sum_{k_1\mydots k_{M}}\sum_{ l_1 \mydots l_{M}} \left(\prod_{i=1}^M \varepsilon^{k_i}\right) A_{l_i}^{k_i}(i)\rho_s(t_{0})\nonumber\\ 
&\qquad\times\sum_{m=1}^{M} \mathrm{tr}_B(B^{(1)}_{l_{M+1}}(M+1)U^B_{M,m+1}B^{(1)}_{l_m}(m)\rho_B) \delta_{k_m,1} \nonumber\\
&\qquad\qquad I(k_1l_1,\mydots, k_{m-1} l_{m-1},0 0,k_{m+1} l_{m+1}\mydots, k_Ml_M),  \nonumber
\end{align}
including only the terms at most linear in $\varepsilon$.
Collecting the terms linear in $\varepsilon$ yields the derivative of $\rho_s$\cite{originalheom}: 
\begin{align} \label{eq:sysdgl}
&\partial_t \rho_s(t)=-\frac{i}{\hbar} H_{s,-}(t)\rho_s(t) \\
&\quad+\sum_{l \tilde{l}} A_l^{(1)}(t) \int_{t_0}^{t} \mathrm{d}t_1 \langle B^{(1)}_l(t)B_{\tilde{l}}^{(1)}(t_1)\rangle_B  \rho_{s \tilde{l}}^{(1)}(t,t_1),\nonumber
\end{align} where $\langle A\rangle_B= \mathrm{tr}_B(A\rho_B)$ and the bath correlation function is in the interaction picture, and \red{the first order auxiliary density matrix (ADM) \(\rho^{(1)}\) reads}
\begin{align} 
&\rho_{s l}^{(1)}(t,\tilde{t})= \delta_{k_m 1} \delta_{l_m l} \left\langle T_\leftarrow \prod^{M}_{i=1. i\neq m} B_{l_i}^{(k_i)}(i) \right\rangle_B \nonumber\\ &\qquad \sum_{k_1\dots k_M} \sum_{ l_1\dots l_M} T_{\leftarrow}  A_{l}^{(1)}(m) \left(\prod_{i=1, i\neq m}^M \varepsilon^{k_i} A_{l_i}^{(k_i)}(i) \right)\rho_s(t_0) \nonumber  ,
\end{align}
with $\tilde{t}=m \varepsilon+t_0$.
Here, the derivation deviates from the original recipe of Kubo and Tanimura, since the assumption of a spectral density in Debye form (simple exponential $e^{-\gamma t}$ in time) is not compatible with systems including delay. Generalizations of HEOM usually rely on a decomposition of the spectral density into a sum of exponential functions to recover the Debye form. However, an expansion of the correlation function for the delay case using  $e^{-\gamma |t-t_{delay}|}$ does not yield the advantages of Kubo's and Tanimura's approach, since the original relies on a time constant derivative of the Debye spectral density time correlation function. Instead, a delayed correlation of the above form introduces a sign change at \(t=t_{delay}\), so that a dependence of \( \rho^{(n)}\) on earlier integration times is unavoidable in the case with delay. Thus the integration over $t_1$ is not included in the definition of $\rho^{(1)}$ in contrast to the original HEOM \cite{originalheom}. Keeping the general form of the bath correlation function is more flexible than using a special form, which would simplify the equations of motion in the following. $\rho_{s l}^{(1)}(\cdot,t_1)$ describes bath disturbances to the system density matrix, which are initially caused by an interaction with $ A_{l}^{(1)} $ at time $t_1$ (similar to the auxiliary dimensions in extended TCL \cite{RICHTER2010711}). Of course, the additional time argument prevents direct numerical implementations for increasing $n$. But specific bath correlation functions together with analytic calculation or tensor network methods \cite{orus2014practical,schollwock2011density,CIRAC2017100,verstraete2006matrix,vidal2007classical,CIRAC2017100,plenioheisenberg,PhysRevLett.116.237201,Rosenbach_2016,PhysRevLett.116.093601,kaestle2020protected,schroder2019tensor,PhysRevLett.123.100502} will allow solutions nevertheless.
Using the same technique as for $\partial_t\rho_s(t)$ yields:
\begin{align} \label{eq:dglrho1}
&\partial_t  \rho_{s l_1}^{(1)}(t,t_1)=-\frac{i}{\hbar} H_{s,-}(t)\rho_{s l_1}^{(1)}(t,t_1) \nonumber\\
&\quad +\sum_{l_2\tilde{l}_2} A_{l_2}^{(1)}(t) \int_{t_0}^{t} \mathrm{d}t_2  \langle B^{(1)}_{l_2}(t)B_{\tilde{l}_2}^{(1)}(t_2)\rangle_B  \rho_{s l_1 \tilde{l}_2}^{(2)}(t,t_2,t_1) \nonumber\\
&\quad+\delta(t-t_1)A_{l_1}^{(1)}(t_1) \rho_s(t_1-0^+).
\end{align}
where we use the interaction picture for the bath correlation function.
Instead of an initial condition  $\rho_{s l_1}^{(1)}(t_1,t_1)=A_{l_1}(t_1) \rho_s(t_1-0^+)$, the $\delta$ term at the time of the initial condition is included, i.e., $\rho_{s l_1}^{(1)}(\cdot,t_1)$ is equal to zero (in the delta case) or not defined (in the initial condition case) before time $t_1$. Note that $t_1$, $t_2$ of \red{the second order ADM} $\rho^{(2)}(t,t_2,t_1)$ are not time ordered since different delay/retardation times can occur in open quantum systems.

The form of $\rho^{(2)}$ points to a general definition of \red{the n-th order ADM} $\rho^{(n)}$ starting with $\rho^{(0)}(t)=\rho_s(t)$:
\begin{align}
&\rho_{s \tilde{l}_1\mydots \tilde{l}_n}^{(n)}(t,\tilde{t}_n,\mydots,\tilde{t}_1)=\nonumber\\ &\qquad \qquad
\sum_{k_1\dots k_M} \sum_{ l_1\dots l_M}T_{\leftarrow}
\left(\prod_{j=1}^n A_{\tilde{l}_j}^{(1)}(m_j)\delta_{\tilde{l}_j,l_{m_j}} \delta_{\tilde{k}_{m_j}1}\right) \nonumber\\
&\qquad\qquad\quad\left(\prod_{j=1, \wedge_{i=1}^n j\neq m_i}^M \varepsilon^{k_j} A_{l_j}^{(k_j)}(j) \right)\rho_s(t_0) \nonumber\\ &\qquad\qquad\qquad \left\langle T_\leftarrow \prod^{M}_{j=1, \wedge_{i=1}^n j\neq m_i} B_{l_j}^{(k_j)}(j)\right\rangle_B,
\end{align}
with $\tilde{t}_i=m_i \varepsilon+t_0$.
Analogous to $\rho^{(1)}$, this yields:
\begin{align}
&\partial_t  \rho_{s l_1\mydots l_n}^{(n)}(t,t_1, \mydots, t_{n})=-\frac{i}{\hbar} H_{s,-}(t)\rho_{s l_1\mydots l_n}^{(n)}(t,t_1, \mydots, t_{n})\nonumber\\
&+\sum_{l_{n+1}\tilde{l}_{n+1}} A_{l_{n+1}}^{(1)}(t) \int_{t_0}^{t} \mathrm{d}t_{n+1}  \langle B^{(1)}_{l_{n+1}}(t)B_{\tilde{l}_{n+1}}^{(1)}(t_{n+1})\rangle_B \nonumber\\ &\qquad \qquad\rho_{s l_1 \dots l_n \tilde{l}_{n+1}}^{(n+1)}(t,t_1,\dots,t_{n+1})\nonumber\\ 
&+\sum_{p=1}^n\delta(t-t_{p})A_{l_p}^{(1)}(t_p)\label{gen_eqmo}\\ & \quad \times\rho_{s l_1\mydots l_{p-1}l_{p+1} \mydots l_n}^{(n-1)}(t_p-0^+,t_1, \mydots, t_{p-1},t_{p+1},\mydots t_{n+1}). \nonumber
\end{align}
The last term is again a replacement to an initial condition:  $\rho_{s l_1\mydots l_n}^{(n)}(t_p, t_1, \mydots, t_{n})=A_{l_p}(t_p)\rho_{s l_1\mydots l_{p-1}l_{p+1} l_n}^{(n-1)}(t_p - 0^+,t_1, \mydots, t_{p-1},t_{p+1}, \mydots,t_{n})$ with $t_p=\mathrm{max}_i(t_i)$, and it is clear that $\rho_{s l_1\mydots l_n}^{(n)}(t, t_1, \mydots, t_{n})=0$ for $t < t_{p}$. So for the last term only $p$ with the largest time $t_p$ contributes. Furthermore, \red{the ADM} $\rho^{(n)}$ is invariant under permutations of $t_1, \mydots, t_n$ including their corresponding $l_1, \mydots, l_n$.\\
\red{The HEOM analog scales exponentially with \(n\) in both the number of indices \(l_i\), which contain the possible states involved in the initial interaction at time \(t_i\), as well as in the number of additional time arguments \(t_i\). The number of possible initial states per \(l_i\) and the number of necessary timesteps per \(t_i\) enter into the base of this exponential scaling. For specific applications (e.g., the delta-like correlation functions used in the example below), the number of necessary timesteps can be significantly reduced to include only a short timeframe (e.g., the delay time).}\\
The physics behind Eq.~\eqref{gen_eqmo} is very accessible:
\red{Under the rotating wave approximation and for a bath with negligible initial photon number,} $n$ corresponds to the maximum number of photons \red{propagating between two systems through the bath} at a given time $t$, so an exact truncation of the equations based on the traveling photons is possible. \red{For cases where these assumptions do not hold, such an intuitive physical interpretation of the ADMs is not possible.}
Note, the photons on the left and right side states of the density matrix count accumulating, so a transfer of a single photon density requires two traveling photons (left and right side of density matrix), as opposed to one traveling photon for a single photon coherence.
For other open quantum system equations of motion techniques such as Nakajima-Zwanzig \cite{breuer2002theory} or time convolution less (TCL) equations \cite{breuer2002theory}, the generators $\mathcal{K}$ in the equations of motion contain the system-bath coupling in any order. A calculation of higher-order contributions from $\mathcal{K}$ is generally cumbersome involving higher products of system-bath correlation functions as well as a truncation at a given photon number.
For the HEOM analog, only one system-bath correlation function appears in the second term of Eq.~\eqref{gen_eqmo} cleanly separating on photon number. The first term of Eq.~\eqref{gen_eqmo} describes the system dynamics. The second term represents the absorption of a bath photon, which entered the bath at time $t_{n+1}$. The last term describes photon emission into the bath.

\section{Application to photon propagation}
As a benchmark for the new approach, we consider two spatially separated quasinormal mode (QNM) cavities, coupled to a common photonic bath (Fig.~\ref{fig:model}(a)). The QNMs \(\tilde{f}_{\mu}\) are an open system analog to normal modes, which solve the Helmholtz equation under an outgoing radiation condition \cite{garcia1976resonant, lee1999dyadic, muljarov2011brillouin, kristensen2012generalized, sauvan2013theory, franke2019quantization, kristensen2020modeling}. QNMs have complex eigenfrequencies \(\tilde{\omega}_{\mu} = \omega_{\mu}-i\gamma_{\mu}\) with photon decay rate $\gamma_{\mu}>0$. Here, two dielectric slabs serve as QNM cavities as in Fig.~\ref{fig:model}(b). We assume an effective 1D problem with homogenous continuation in the \(y,z\) direction. The model allows the analytical calculation of the modes (assuming a constant real permittivity \(\epsilon_R\)) and coupling elements (cf.~\apprefA). We include only the lowest energy QNM, assuming that all other modes are off-resonance. Since the slabs are identical, both have the same frequency \(\tilde{\omega}_A = \tilde{\omega}_B = \tilde{\omega}_1\). However, we keep the indices for generality. The slabs are separated by the distance \(R\), which is large enough for a separate quantization of the modes without direct inter-cavity coupling. \red{Instead, the QNMs couple to a common surrounding bath. This interaction is described by the Hamiltonian}
\begin{align}\label{eq:sysbathhamiltonian}
    H_{SB} = \hbar\sum_{\mu=A,B}\int\mathrm{d}x\int_0^{\infty}\mathrm{d}\omega g_{\mu}(x,\omega)\hat{c}(x,\omega)\hat{a}_{\mu}^{\dagger} + \mathrm{H.a.},
\end{align}
\red{where \(\hat{a}_{\mu}\) are the QNM operators for slab \(\mu\). The bath operators \(\hat{c}(x,\omega)\) are assumed to be Bosonic. The derivation of the Hamiltonian and coupling elements \(g_{\mu}(x,\omega)\) are shown in \apprefA}.

\subsection{Equations of motion for two traveling photons}
\label{EOM}
As a first step, we limit the dynamics to cases with at most two propagating photons (one on each side or two on one side of the density matrix). Therefore, the hierarchy truncates at \red{the second order ADM}, i.e., \(\rho^{(n)} = 0, n>2\), and:
\begin{align} \label{eq:rho2sol}
    &\rho^{(2)}_{s,l_1,l_2}(t, t_1, t_2) =\nonumber \\ 
    &\quad\Theta(t_1-t_2)U^s(t,t_1)A^{(1)}_{l_1}(t_1)\rho^{(1)}_{s, l_2}(t_1-0^+,t_2)\nonumber \\
    &\quad+\Theta(t_2-t_1)U^s(t,t_2)A^{(1)}_{l_2}(t_2)\rho^{(1)}_{s,l_1}(t_2-0^+,t_1), 
\end{align}
using the initial conditions for \(\rho^{(2)}\). Inserting Eq.~\eqref{eq:rho2sol} into Eq.~\eqref{eq:dglrho1}, we obtain:
\begin{align} \label{eq:dglrho1closed}
    &\partial_t  \rho_{s l_1}^{(1)}(t,t_1)=-\frac{i}{\hbar} H_{s,-}(t)\rho_{s l_1}^{(1)}(t,t_1) \nonumber\\
    & \qquad +\sum_{l_2\tilde{l}_2} A^{(1)}_{\tilde{l}_2}(t) \int_{t_0}^{t_1} \mathrm{d}t_2  \langle B^{(1)}_{\tilde{l}_2}(t)B_{l_2}^{(1)}(t_2)\rangle_B \nonumber \\
    & \qquad\qquad\qquad \times U^s(t,t_1)A^{(1)}_{l_1}(t_1)\rho^{(1)}_{s l_2}(t_1-0^+,t_2) \nonumber\\
    &\qquad +\sum_{l_2\tilde{l}_2} A^{(1)}_{\tilde{l}_2}(t) \int_{t_1}^{t} \mathrm{d}t_2  \langle B^{(1)}_{\tilde{l}_2}(t)B_{l_2}^{(1)}(t_2)\rangle_B  \nonumber\\
    & \qquad\qquad\qquad \times U^s(t,t_2)A^{(1)}_{l_2}(t_2)\rho^{(1)}_{s l_1}(t_2-0^+,t_1) \nonumber \\
    &\qquad+\delta(t-t_1)A^{(1)}_{l_1}(t_1) \rho_s(t_1-0^+) .   
\end{align}
Eqs.~\eqref{eq:dglrho1closed} and \eqref{eq:sysdgl} form a closed set of equations of motion for the system density matrix that are exactly solvable (cf.~\apprefB) for at most two traveling photons.
Fig.~\ref{fig:model}(c) illustrates connections between the equations with one photon traveling from time \(t_1=t-\tau\) until \(t\) through the bath, requiring the calculation of \(\rho^{(1)}\). Intermittently a second photon is emitted into the bath at \(t_2\). 

\begin{figure}
    \centering
    \includegraphics[width = 8.6cm]{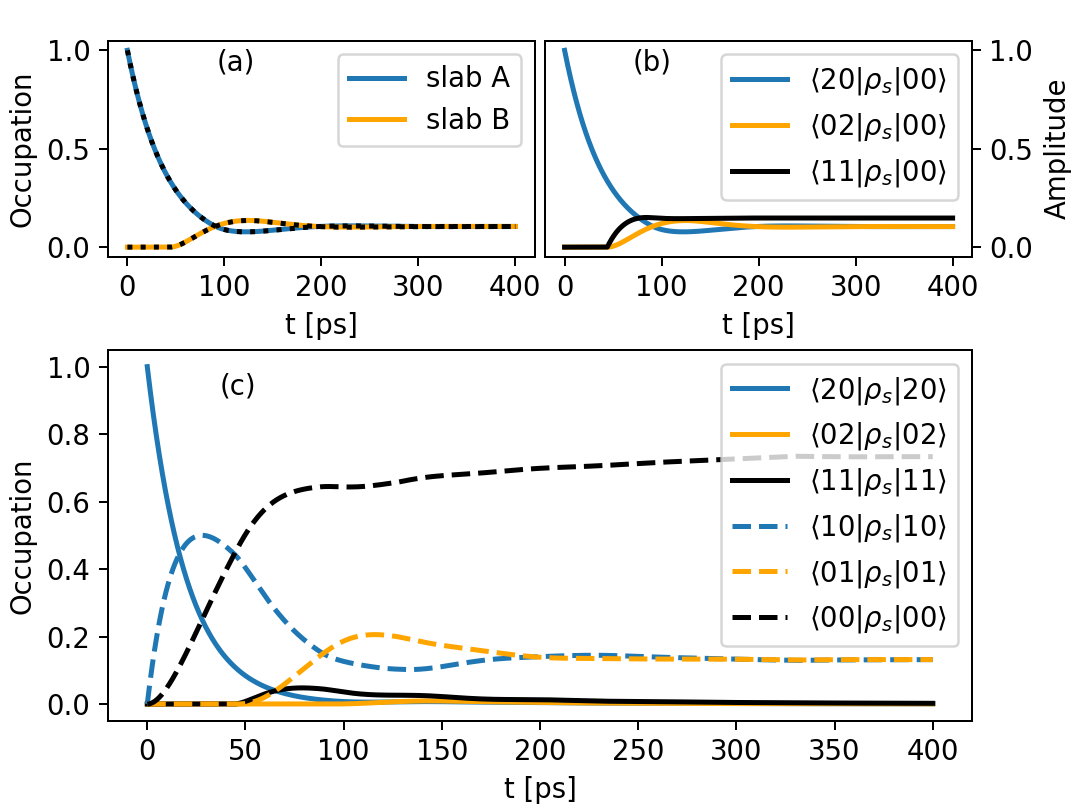}
    \caption{Dynamics of photon exchange between two dielectric slabs. (a) Single-photon occupations in the slabs for an initial state with one photon in slab A. The dotted lines show the full wave function solution. (b) Two-photon coherences. (c) Approximate dynamics of the occupations with initially two excitations in slab A. In all cases, the QNM frequencies of the slabs are identical \(\tilde{\omega}_1 = (0.06 -0.0124i)\,\mathrm{eV}\), with coupling strength \(V_{BA} = V_{AB} = 0.0062\,\mathrm{eV}\), and delay time \(\tau\approx 44\,\mathrm{ps}\).}
    \label{fig:plots}
\end{figure}

The dynamics of a specific system are determined by the system-bath correlation function \(\langle B^{(1)}_{\tilde{l}_2}(t)B_{l_2}^{(1)}(t_1)\rangle_B\), which describes the emission of a photon into the bath at time \(t_1\) and reabsorption at time \(t\). \red{For applications, the abstract operators \(B_l^{(1)}\) have to be replaced with operators adapted to the problem. Comparing the system-bath Hamiltonian from Eq.~\eqref{eq:sysbathhamiltonian} to the abstract form suggests the replacements}
\begin{align} \label{eq:opreplace}
    &A_l(t) \to \hat{A}^{\alpha}_{\nu_1\nu_2}(t),\nonumber\\
    &B_l(t) \to\\
    &\quad\sum_{\mu = A,B}\langle \nu_1| \hat{a}_{\mu}|\nu_2\rangle\int\mathrm{d}x\int_0^{\infty}\mathrm{d}\omega g^{*}_{\mu}(x,\omega)\hat{c}^{\dagger\alpha}(x,\omega)+\mathrm{H.a.},\nonumber
\end{align}
\red{where \(\hat{A}_{\nu_1\nu_2} = |\nu_1\rangle\langle\nu_2|\) with system states \(|\nu_i\rangle\), and \(\alpha = L,R\) for left/right Liouville space operators. The resulting correlation function thus reads}
\begin{align*}
    \langle B^{(1)}_{\tilde{l}_2}(t)B_{l_2}^{(1)}(t_1)\rangle_B \to C^{\alpha}_{\nu_1\nu_2\nu_3\nu_4}(t,t_1),
\end{align*}
\red{with \(C^{L}_{\nu_1\nu_2\nu_3\nu_4}(t,t_1) = \sum_{\mu\eta}\langle \nu_1| \hat{a}^{\dagger}_{\mu}|\nu_2\rangle\langle \nu_3| \hat{a}_{\eta}|\nu_4\rangle C_{\mu\eta}(t-t_1)\) and \(C^{R}_{\nu_1\nu_2\nu_3\nu_4}(t,t_1)= \left(C^{L}_{\nu_1\nu_2\nu_3\nu_4}(t,t_1)\right)^*\). The index \(\alpha\) refers to the interaction at time \(t_1\). The QNM correlation function \(C_{\mu\eta}\) for the two coupled dielectric slabs from Fig.~\ref{fig:model}(b) reads} (cf. \apprefA)
\begin{align}
    C_{\mu\eta}(t-t')\approx 2V_{\mu\eta}\hbar^2&\left(\Theta(t-t')\delta(t-t'-\tau)\right.\nonumber\\
    &+\left.\Theta(t'-t)\delta(t-t'+\tau)\right) \label{eq:corrfunc}.
\end{align}
The coupling strength is given by \(V_{\mu\eta} = (1+\delta_{\mu\eta})\gamma_1/2\) with the cavity decay rate \(\gamma_1\). Due to the topology of the system, the inter-cavity coupling is exactly half the dissipation rate. For the 1D case, a photon emitted away from the other cavity will not return, while a photon emitted towards the other cavity can be transferred into that cavity. In higher dimensions, the inter-cavity coupling will generally be much smaller than the dissipation rate. The delay time \(\tau\) in Eq.~\eqref{eq:corrfunc} depends implicitly on the involved cavities, with \(\tau=(1-\delta_{\mu\eta})R/c\).

\subsection{Exact inter-cavity dynamics using HEOM}
For one initial excitation (one photon on each side of the density matrix), three system states \(|A\rangle = |10\rangle,\, |B\rangle = |01\rangle,\, |0\rangle = |00\rangle\) contribute, with the excitation in slab A or  B, or both slabs in the ground state, respectively. For this setup, {\it the truncation of the HEOM is exact} since the maximal number of propagating photons at any time is set by the initial conditions.
Inserting Eq.~\eqref{eq:corrfunc} into Eq.~\eqref{eq:sysdgl} yields the equations of motion. As an example, the occupation in slab A \(\langle A|\rho_s(t)|A\rangle\) evolves as (cf.~\apprefB):
\begin{align} \label{eq:rho0example} 
     &\partial_t \langle A|\rho_s(t)|A\rangle = -2\gamma_A  \langle A|\rho_{s}(t)|A\rangle \nonumber\\
     &\qquad\qquad-2V^*_{BA}\mathrm{e}^{i\omega_B\tau} \langle 0|\rho_{s, 0B}^{(1)L}(t,t-\tau)|A\rangle +\mathrm{c.c.}
\end{align}
For the auxiliary density matrix \(\rho^{(1)}\), starting from Eq.~\eqref{eq:dglrho1closed} and switching to a rotating frame (cf.~\apprefB) results in:
\begin{align}\label{eq:rho1example}
    \partial_{t}\langle 0|\rho_{s, 0B}^{(1)L}&(t,t_1)|A\rangle = \delta(t-t_1)\langle B|\rho_s(t_1)|A\rangle\nonumber\\
    &  -\gamma_A\langle 0|\rho_{s, 0B}^{(1)L}(t,t_1)|A\rangle\nonumber\\
    &-2V_{BA}\mathrm{e}^{-i\omega_B\tau}\langle B|\rho_{s, B0}^{(1)R}(t_1,t-\tau)|0\rangle\nonumber\\
    &-2V_{BA}\mathrm{e}^{-i\omega_A\tau}\langle 0|\rho_{s, 0B}^{(1)L}(t-\tau,t_1)|A\rangle.
\end{align}
The remaining equations for the occupation in B, the coherences, and matrix elements for \(\rho^{(1)}\) are of a similar form (cf.~\apprefB). Time-local processes such as cavity photon dissipation are included in the zeroth step of the hierarchy. For time-non-local interactions, the system density matrix in Eq.~\eqref{eq:rho0example} only couples to the first auxiliary density matrix \(\rho^{(1)}\). 

Fig.~\ref{fig:plots}(a) shows the time dynamics of the single-photon occupations in slabs A and B. The model system allows a calculation using the wave function (cf.~\apprefC) as a benchmark.
The HEOM (solid lines) and exact wave function (dotted lines) results agree perfectly. Over time the single excitation in slab A will dissipate into the bath. However, some photons are transferred to the QNM of slab B with delay \(\tau\approx 44\, \mathrm{ps}\).  For the used parameters, the occupation in B is even larger than the occupation in A after some time. Eventually, the system arrives at a trapped state \cite{PhysRevLett.116.093601, bello2019unconventional, hughes2017anisotropy, grimsmo2015time, nemet2019comparison, barkemeyer2020revisiting, finsterholzl2020nonequilibrium} due to interference from the inter-cavity transfer.

Note that the HEOM allows in principle the inclusion of Lindblad terms (e.g. for pumping), which the wave function does not. Also, an extension to two-photon processes is feasible for the HEOM.
Fig.~\ref{fig:plots}(b) shows the two-photon coherences (two photons on one side of the density matrix, none on the other) for the two slabs from Fig.~\ref{fig:model}(b), which includes  at most two traveling photons, resulting in a calculation analogous to Fig.~\ref{fig:plots}(a)(cf.~\apprefB). The amplitudes of the intra-cavity coherences \(\langle 20|\rho_s|00\rangle\)/\(\langle 02|\rho_s|00\rangle\) resemble the dynamics of the densities in Fig.~\ref{fig:plots}(a), since in principle the same independent processes are involved. The inter-cavity coherence \(\langle 11|\rho_s|00\rangle\) requires the transfer of just one photon and thus shows a rapid increase after \(t=\tau\). In the final equilibrium state, the probability (coherence squared) of the inter-cavity contribution matches the sum of the two intra-cavity probabilities.

\subsection{Approximate truncation of multi-photon processes}
A feasible calculation of the exact solution as shown here is limited to a small number of photons by the exponential scaling of the numerical complexity with the number of excitations. For systems requiring a higher number of traveling photons, a calculation of the higher steps in the hierarchy via matrix product states or other tensor networks \cite{orus2014practical,schollwock2011density,CIRAC2017100,verstraete2006matrix,vidal2007classical,CIRAC2017100,plenioheisenberg,PhysRevLett.116.237201,Rosenbach_2016,PhysRevLett.116.093601,kaestle2020protected,schroder2019tensor,PhysRevLett.123.100502}  may be possible as well as analytic calculations in special setups.\\
The HEOM also allows a perturbative truncation of the hierarchy for systems with a small system-bath coupling. Thus, at least an approximate solution is possible for higher excitation numbers. Such an approximative solution is shown in Fig.~\ref{fig:plots}(c) for the case of the two slabs with an initial population of two  excitations in slab A. In principle, this setup can show up to four propagating photons (two on each side of the density matrix). For small inter-cavity couplings \(V_{\mu\eta}\), however, the timescale on which photons are exchanged between the cavities is longer than the propagation time \(\tau\) of the photons. Therefore assuming at most two photons traveling through the bath at any time, a truncation of the hierarchy at the second step, i.e., \(\rho^{(n)}=0\) for \(n>2\), may give good results. Under this assumption, the equations of motion reduce to the closed set of equations from Sec.~\ref{EOM}, and the dynamics are calculated in the same way as for the one-photon densities (cf. \apprefB). Here, the two-photon population in slab A \(\langle 20|\rho_s|20\rangle\) decays exponentially while emitting photons into the bath. Because of this instant emission, the occupation \(\langle 10|\rho_s|10\rangle\) with one photon in slab A and the ground state \(\langle 00|\rho_s|00\rangle\) with no photons in the cavities increase immediately. In contrast, the states \(\langle 11|\rho_s|11\rangle\) with one photon in each slab and \(\langle 01|\rho_s|01\rangle\) with one photon in slab B only increase after \(t=\tau\), since the photons need to travel \(R=c\tau\) between the slabs. However, the density \(\langle 02|\rho_s|02\rangle\)  only increases after \(t=2\tau\). This is an artifact of the two-photon truncation of the HEOM since the transfer from \(\langle 20|\rho_s|20\rangle\) to \(\langle 02|\rho_s|02\rangle\) requires four propagating photons (two on each side). \red{In contrast, the transfer of the two-photon coherence from Fig.~2(b) requires only two photons on one side, so that the coherence \(\langle 02|\rho_s|00\rangle\) increases already after \(t=\tau\), even though the density \(\langle 02|\rho_s|02\rangle\) takes twice as long to increase. If the photon transfer rate and delay time are small enough, this error is expected to be small, if few enough photons are transferred at once during the time \(\tau\). However, a small, qualitative difference to the exact solution is unavoidable.} For larger coupling strengths or delay times, the approximate solution will deviate increasingly from the real solution and additional steps in the hierarchy must be included.

\section{Conclusion}

In conclusion, we analyzed the derivation of hierarchical equations of motion and transferred the idea to open quantum systems with delay.
The resulting equations allow a natural, easy truncation on the number of excitations in the bath, which is otherwise cumbersome for Nakajima-Zwanzig or time convolution-less equations.
The first implementation for single- and multi-photon transfer between two cavities demonstrated the feasibility of the approach.
We expect that in the future more demanding implementations including tensor network approaches may allow the simulation of several photons traveling through complex quantum networks.



\appendix

\section{Analytic coupling elements}
\label{app1}
We use analytic expressions of the mode frequencies, decay constants, and coupling elements for numeric evaluation. For linearly polarized waves and assuming a homogeneous continuation in \(y,z\)-direction, the problem reduces to the 1D model from Fig.~\ref{fig:model}(b). The QNM within each slab is given by \cite{lalanne2018light, kristensen2020modeling} 
\begin{align}
    \left.\tilde{f}_{\mu}(x)\right|_{|x|<L/2} = \mathrm{e}^{in_Rk_{\mu}x}+\mathrm{e}^{-in_Rk_{\mu}x+i\mu\pi},
\end{align}
where \(n_R = \sqrt{\epsilon_R}\) is the refractive index of the slab and \(k_{\mu} = \tilde{\omega}_{\mu}/c\) is the QNM wavenumber. The QNM frequency \(\tilde{\omega}_{\mu}\) is \cite{lalanne2018light, kristensen2020modeling}
\begin{align}
    \tilde{\omega}_{\mu}L/c = \frac{2\pi\mu+i\mathrm{ln}\left((n_R-n_B)^2/(n_r+n_B)^2\right)}{2n_R}.
\end{align}
Thus, the frequency of the first QNM \(\tilde{f}_1(x)\) is \(\tilde{\omega}_1 = \omega_1 -i\gamma_1= (1-i0.21)L/c\). The second QNM \(\tilde{f}_2(x)\) has a resonance frequency that is twice as large. Hence, as a first approximation, we take only the first QNM in our calculations.\\
Outside of the cavity (\(|x|>L/2\)), we replace the QNMs with regularized modes \cite{ge2014quasinormal} \(\tilde{F}_{\mu}(x,\omega) = \int_{-L/2}^{L/2}\mathrm{d}x' G_B(x,x',\omega)\Delta\epsilon(x')\tilde{f}_{\mu}(x') = (x/|x|)M_{\mu}(\omega)\mathrm{e}^{i\omega|x|/c}\), where \(\Delta\epsilon(x) = \epsilon_R-\epsilon_B, |x|<L/2\), and \(0\) otherwise, and 
\begin{align}
    M_{\mu}(\omega) = \frac{i}{2}L(\pi^2-1)&\left[\mathrm{si}\left(\frac{(\omega+\pi\tilde{\omega}_{\mu})L}{2c}\right)\right.\nonumber\\
    &\left.-\mathrm{si}\left(\frac{(\omega-\pi\tilde{\omega}_{\mu})L}{2c}\right)\right]
\end{align}
is an analytical factor that vanishes for \(\omega\to\infty\). \(\mathrm{si}(x) = \mathrm{sin}(x)/x\) is the unnormalized sinc-function. \(G_B(x,x',\omega) = i\mathrm{e}^{-i\omega|x-x'|/c}/2\) is the vacuum Green's function for the case of linearly polarized waves, solving the Helmholtz equation 
\begin{align} \label{eq:backgreen}
    \left(\partial_x^2 +\frac{\omega^2}{c^2}\right)G_B(x,x',\omega) = \frac{\omega^2}{c^2}\delta(x-x').
\end{align}
We locate the slab A at \(x=0\) and slab B at \(x=R\) (cf. Fig.~\ref{fig:model}(b)), so that \(\tilde{f}_1(x) = \tilde{f}_A(x)\) and \(\tilde{f}_B(x) = \tilde{f}_A(x-R)\). \\
We quantize the QNMs following the procedure laid out in \cite{franke2019quantization}, with minor adjustments due to the 1D nature of the problem, e.g., taking the 1D analog of the electric field quantization and QNM Green's function instead of the 3D expressions that were used in  \cite{franke2019quantization}. Since the QNM quantization relies on a complex permittivity, we add  a constant imaginary part to the permittivities of the slabs and background medium:  \(\epsilon^{\alpha} = \epsilon_{R/B} + i\alpha\kappa\) (cf. \cite{franke2020fluctuation}) so that the original values are retained in the limit \(\alpha\to 0\). Taking the 1D analog of the quantization in dissipative media from \cite{PhysRevA.53.1818}, we find the electric field operator to be 
\begin{align} \label{eq:efeldvogelwelsch}
    E^{\alpha}(x) = \int_0^{\infty}\mathrm{d}\omega\int\mathrm{d}x' \frac{i}{\omega\epsilon_0}G^{\alpha}(x,x',\omega)\hat{j}^{\alpha}(x',\omega)+\mathrm{H.a.},
\end{align}
where \(G(x,x',\omega)\) is the Greens function of the dissipative medium and \(\hat{j}^{\alpha}(x,\omega) = \omega\sqrt{(\hbar\epsilon_0/\pi)\epsilon^{\alpha}_I(x,\omega)}\hat{b}(x,\omega)\) is the noise-current density operator, with \(\hat{b}(x,\omega)\) a Bosonic photon annihilation operator. \(\epsilon_I\) is the imaginary part of the permittivity, which is frequency independent in the model from Fig.~\ref{fig:model}(b), but we keep the frequency dependence for generality.  We use the Green's function expansion in terms of QNMs \cite{lee1999dyadic,ge2014quasinormal, lalanne2018light} \(G(x,x',\omega) = \sum_{\mu = A,B}A_{\mu}(\omega)\tilde{f}_{\mu}(x)\tilde{f}_{\mu}(x')\), where \(A_{\mu}(\omega) = \omega/(2(\tilde{\omega}_{\mu}-\omega))\), and the QNM functions \(\Tilde{f}_{\mu}\) are replaced with regularized modes $\tilde{F}_{\mu}$ outside their respective cavity volumes. Inserting the QNM Green's function into Eq.~\eqref{eq:efeldvogelwelsch}, we find QNM operators analogous to \cite{franke2019quantization}:
\begin{align} \label{eq:nonsymmops}
    &\tilde{a}_A = \sqrt{\frac{2}{\pi\omega_A}}\int_0^{\infty}\mathrm{d}\omega A_A(\omega)\nonumber\\&\qquad\times\left[ \int_{-L/2}^{L/2}\mathrm{d}x \sqrt{\epsilon^{\alpha}_I(x,\omega)}\tilde{f}_A^{\alpha}(x)\hat{b}(x,\omega)\right.\nonumber\\
    &\qquad\quad+\lim_{\lambda\to\infty}\int_{L/2}^{\lambda}\mathrm{d}x \sqrt{\epsilon^{\alpha}_I(x,\omega)}\tilde{F}_A^{\alpha}(x,\omega)\hat{b}(x,\omega)\nonumber\\
      &\qquad\quad+\left.\lim_{\lambda\to\infty}\int_{-\lambda}^{-L/2}\mathrm{d}x \sqrt{\epsilon^{\alpha}_I(x,\omega)}\tilde{F}_A^{\alpha}(x,\omega)\hat{b}(x,\omega)\right],
\end{align}
which depend implicitly on \(\alpha\to 0\). In the first integral, the limit \(\alpha\to 0\) can be carried out immediately, so that this contribution vanishes, because \(\lim_{\alpha\to 0}\epsilon^{\alpha}_I = 0\). In the other two integrals, the order of the limits cannot be exchanged, as pointed out in \cite{franke2020fluctuation}, so the limit \(\lambda\to\infty\) has to be taken first. The operators for the QNMs of cavity B are defined analogously, just spatially shifted by \(R\).\\
The QNM operators defined in Eq.~\eqref{eq:nonsymmops} are non-Bosonic, with \(\left[\tilde{a}_A,\tilde{a}_A^{\dagger}\right] = S_{AA}\), and
\begin{align}
    S_{AA} =& \frac{2}{\pi\omega_A}\int_0^{\infty}\mathrm{d}\omega A_A(\omega)A^*_A(\omega)\nonumber\\
    &\times\left[\lim_{\lambda\to\infty}\int_{L/2}^{\lambda}\mathrm{d}x \epsilon^{\alpha}_I(x,\omega)\tilde{F}_A^{\alpha}(x,\omega)\tilde{F}_A^{*,\alpha}(x,\omega)\right.\nonumber\\
     &\quad+\left.\lim_{\lambda\to\infty}\int_{-\lambda}^{-L/2}\mathrm{d}x \epsilon^{\alpha}_I(x,\omega)\tilde{F}_A^{\alpha}(x,\omega)\tilde{F}_A^{*,\alpha}(x,\omega)\right].
\end{align}
Analogous to \cite{franke2020fluctuation}, we employ the Helmholtz equation of the background Green's function (Eq.~\eqref{eq:backgreen}) to reduce the integral over \(x\) to the value of the modes at the limits of the integration volume. Taking the limit \(\lambda\to\infty\) first and then \(\alpha\to 0\), we find
\begin{align}\label{eq:SAA}
    S_{AA} = \frac{2c}{\gamma_1}\left|M_1(\tilde{\omega}_1)\right|^2,
\end{align}
where we used \(\tilde{\omega}_A =  \tilde{\omega}_1\).\\
The overlap integral \([\tilde{a}_A,\tilde{a}_B^{\dagger}] = S_{AB}\) is calculated accordingly. We make use of the fact that the two slabs are identical except for their spatial separation and hence \(\tilde{\omega}_A = \tilde{\omega}_B = \tilde{\omega}_1\), to obtain
\begin{align}\label{eq:SAB}
    S_{AB} = \frac{2c}{\gamma_1}\left|M_1(\tilde{\omega}_1)\right|^2\mathrm{Re}\left\{\frac{\tilde{\omega}_1}{2\omega_1}\mathrm{e}^{-i\omega_1 R/c}\right\}\mathrm{e}^{-\gamma_1 R/c}.
\end{align}
Since \(\left|\mathrm{Re}\left\{\tilde{\omega}_1\mathrm{e}^{-i\omega_1 R/c}/(2\omega_1)\right\}\right| < 1\), it follows that \(|S_{AB}/S_{AA}|< \mathrm{e}^{-\gamma_1 R/c}\), due to the retarded interaction between the slabs. The QNMs penetrate through the boundary of the slab so that there is a non-zero overlap even without time delay. However, the mode is concentrated at the cavity so that the overlap is small if the slabs are well enough separated. Below, the correlation functions are discussed  for the case with finite time delay. The QNM wavelength is \(\lambda_1 = 2 L\), so a separation of a few dozen wavelengths, as used in the main text, leads to negligible contributions of the overlap.\\
Thus, the QNM operators are symmetrized independently within their respective cavities similar to the single-cavity case in \cite{franke2019quantization}:
\begin{align}\label{eq:qnmops}
    \hat{a}_{\mu} = \int\mathrm{d}x\int_0^{\infty}\mathrm{d}\omega L_{\mu}(x,\omega)\hat{b}(x,\omega),
\end{align}
with 
\begin{align}
L_{\mu}(x,\omega) = S^{-1/2}_{\mu\mu}\sqrt{\frac{2\epsilon_I(x,\omega)}{\pi \omega_{\mu}}}A_{\mu}(\omega)\tilde{f}_{\mu}(x),
\end{align}
and the mode function \(\Tilde{f}_{\mu}\) is replaced by the regularized mode \(\Tilde{F}_{\mu}\) outside the slab volume. The imaginary part of the permittivity and the bounds of the spatial integral include implicit limits, as discussed above.\\
We now define continuum operators \(\hat{c}(x,\omega) = \hat{b}(x,\omega) - \sum_{\mu=A,B}L_{\mu}^*(x,\omega)\hat{a}_{\mu}\) \cite{franke2020quantized}, which commute with the symmetrized Bosonic QNM operators and serve as the bath. While they are generally non-Bosonic, as a first approximation, we neglect the non-Bosonic contributions. This allows us to decompose the full Hamiltonian \(H = \hbar\int\mathrm{d}x\int_0^{\infty}\mathrm{d}\omega \omega \hat{b}^{\dagger}(x,\omega)\hat{b}(x,\omega)\) into system and bath parts \cite{franke2020quantized}:
\begin{align} \label{eq:QNMHamiltonian}
    H_S &= \hbar \sum_{\mu = A,B}\omega_{\mu} \hat{a}_{\mu}^{\dagger}\hat{a}_{\mu},\nonumber\\
    H_B &= \hbar\int\mathrm{d}x\int_0^{\infty}\mathrm{d}\omega \omega \hat{c}^{\dagger}(x,\omega)\hat{c}(x,\omega),\nonumber \\
    H_{SB} &= \hbar\sum_{\mu=A,B}\int\mathrm{d}x\int_0^{\infty}\mathrm{d}\omega g_{\mu}(x,\omega)\hat{c}(x,\omega)\hat{a}_{\mu}^{\dagger} + \mathrm{H.a.}   
\end{align}
The coupling elements \(g_{\mu}(x,\omega)= -S^{-1/2}_{\mu\mu} \times\sqrt{\epsilon_I(x,\omega)/(2\pi\omega_{\mu})}\omega \tilde{f}_{\mu}(x)\), are derived from the projectors \(L_{\mu}(x,\omega)\), with the pole at \(\omega = \tilde{\omega}_{\mu}\) removed during the derivation, as shown in \cite{franke2020quantized}.\\
To derive the coupling strength of the interaction between the slabs mediated via the bath, we calculate the correlation function \cite{Chernyak:1996, mukamel1999principles} that characterizes the system-bath interaction in the HEOM formalism:
\begin{align} \label{eq:corrfuncgeneral}
    C_{\mu\eta}(t-t') = \hbar^2\int_0^{\infty}\mathrm{d}\omega\int_0^{\infty}\mathrm{d}\omega'\int\mathrm{d}x\int\mathrm{d}x' e^{-i\omega(t-t')}\nonumber\\
    \times g_{\mu}(x,\omega)g^*_{\eta}(x',\omega')\left\langle\hat{c}(x,\omega)\hat{c}^{\dagger}(x',\omega')\right\rangle_B.
\end{align}
For \(\rho_B=|0\rangle\langle 0 |\) (no initial photons), the expectation value results in a delta function, so only an integral over the coupling elements remains. This is calculated similarly to \cite{franke2020quantized}, i.e., by assuming that the coupling is sharply peaked at the QNM frequency, so that
\begin{align}
    \int\mathrm{d}x g_{\mu}(x,\omega)g_{\eta}^*(x,\omega) &\approx S^{-1}_{11}\frac{2c}{\gamma_1}|M_1(\tilde{\omega}_1)|^2 \frac{\gamma_1}{2\pi}\nonumber\\
    &\times\left(\mathrm{e}^{i\omega R_{\mu\eta}/c}+\mathrm{e}^{-i\omega R_{\mu\eta}/c}\right),
\end{align}
where \(|R_{\mu\eta}|\) is R if \(\mu\neq \eta\) and \(0\) otherwise. Using \(S_{11} = 2c|M_1(\tilde{\omega}_1)|^2/\gamma_1\), and defining the retardation time \(\tau = R_{\mu\eta}/c\) as an implicit function of \(\mu\) and \(\eta\), the correlation function becomes
\begin{align}
    &C_{\mu\eta}(t-t') = \frac{\gamma_1\hbar^2}{2\pi}\int_0^{\infty}\mathrm{d}\omega \left(\mathrm{e}^{i\omega \tau}+\mathrm{e}^{-i\omega \tau}\right)\mathrm{e}^{-i\omega(t-t')},
\end{align}
As a final approximation, we extend the lower limit to \(-\infty\) \cite{franke2020quantized}, to obtain the correlation function in Eq.~\eqref{eq:corrfunc}.

\section{Calculation of the equations of motion}
\label{app2}
For equations of motion of the density-matrix elements, \red{we use the replacements from Eq.~\eqref{eq:opreplace} to} convert Eq.~\eqref{eq:sysdgl} to a more explicit form:
\begin{align} \label{eq:rhosysconcr}
    &\partial_t \rho_s(t)=-\frac{i}{\hbar} H_{s,-}(t)\rho_s(t) \nonumber\\
    &\quad+\sum_{\alpha,\beta = L,R}\sum_{\nu_1\dots\nu_4}(-1)^{\alpha+\beta} \hat{A}_{\nu_1\nu_2}^{\alpha}(t)  \nonumber\\
    &\qquad\qquad\times\int_{t_0}^{t} \mathrm{d}t_1 C^{\beta}_{\nu_1\nu_2\nu_3\nu_4}(t,t_1) 
   \rho_{s, \nu_3\nu_4}^{(1)\beta}(t,t_1),
\end{align}
where the sign is negative if \(\alpha\neq\beta\). 
For brevity, we use \(|A\rangle,|B\rangle,|0\rangle\) as defined in the main text, above Eq.~\eqref{eq:rho0example}. To derive Eq.~\eqref{eq:rho0example}, we take the expectation value with respect to state \(|A\rangle\)  on \eqref{eq:rhosysconcr} to obtain,
\begin{align}
    &\partial_t \langle A|\rho_s(t)|A\rangle = -\frac{i}{\hbar}\langle A|H_{s,-}\rho_s(t)|A\rangle\nonumber\\
    &\qquad+\sum_{\alpha,\beta = L,R}\sum_{\nu_1\dots\nu_4}(-1)^{\alpha+\beta}\int_{t_0}^{t} \mathrm{d}t_1 C^{\beta}_{\nu_1\nu_2\nu_3\nu_4}(t,t_1)\nonumber\\
    &\qquad\qquad\qquad\qquad\quad\times\langle A|\hat{A}_{\nu_1\nu_2}^{\alpha}(t)\rho_{s, \nu_3\nu_4}^{(1)\beta}(t,t_1)|A\rangle.
\end{align}
Since the \(|\nu_i\rangle\) are orthogonal, only certain combinations of states and \(\alpha,\beta\) survive. The integral and the sums are eliminated using the definition of the QNM correlation function (Eq.~\eqref{eq:corrfunc}) and the initial conditions for \(\rho^{(1)}\). To avoid fast-rotating terms, we move to a rotating frame, where we use a  rotating-frame representation of \(\rho^{(1)}\) with respect to its time arguments, e.g.,
\begin{align}
    \langle 0 |\rho^{(1),L}_{s,0B}(t,t_1)|A\rangle \to \mathrm{e}^{i\omega_1 (t-t_1)}\langle 0 |\rho^{(1),L}_{s,0B}(t,t_1)|A\rangle,
\end{align}
where we have used \(\omega_A=\omega_B =\omega_1\). This results in the first-order equation of motion given in Eq.~\eqref{eq:rho0example}. Similarly, we obtain an equation for the coherence \(\langle A|\rho_s(t)|B\rangle\):
\begin{align} \label{eq:rho0AB} 
    & \partial_t \langle A|\rho_s(t)|B\rangle = -(\gamma_A+\gamma_B)  \langle A|\rho_{s}(t)|B\rangle \nonumber\\
     &-2V^*_{BA}\mathrm{e}^{i\omega_B\tau} \langle 0|\rho_{s, 0B}^{(1)L}(t,t-\tau)|B\rangle +\mathrm{c.c.}(A\leftrightarrow B).
\end{align}
The equations for the occupation in slab B and the second coherence term are obtained from Eq.~\eqref{eq:rho0example} and Eq.~\eqref{eq:rho0AB}, respectively, by exchanging \(A\leftrightarrow B\).\\ 
For Eq.~\eqref{eq:rho1example}, we insert Eq.~\eqref{eq:opreplace} into Eq.~\eqref{eq:dglrho1closed} and again use the rotating frame. Within the rotating-frame, \(\langle 0 |\rho^{(1),L}_{s,0B}(t,t_1)|A\rangle\) evolves according to Eq.~\eqref{eq:rho1example}. In the same manner, we derive:
\begin{align}\label{eq:rho1AA00}
    \partial_{t}\langle A|\rho_{s, A0}^{(1)R}&(t,t_1)|0\rangle = \delta(t-t_1)\langle A|\rho_s(t_1)|A\rangle\nonumber\\
    &  -\gamma_A\langle A|\rho_{s, A0}^{(1)R}(t,t_1)|0\rangle\nonumber\\
    &-2V^*_{BA}\mathrm{e}^{i\omega_B\tau}\langle 0|\rho_{s, 0B}^{(1)L}(t_1,t-\tau)|A\rangle\nonumber\\
    &-2V^*_{BA}\mathrm{e}^{i\omega_B\tau}\langle B|\rho_{s, A0}^{(1)R}(t-\tau,t_1)|0\rangle.
\end{align}
The last six matrix elements of \(\rho^{(1)}\) are derived from Eq.~\eqref{eq:rho1example} and \eqref{eq:rho1AA00} by complex conjugation or exchanging the indices \(A\) and \(B\). Note that \(\rho^{(1)}(t,t_1)\) vanishes for \(t<t_1\) or \(t_1<0\). Furthermore, only \(\rho^{(1)}(t,t-\tau)\) appears in Eq.~\eqref{eq:rho0example} and Eq.~\eqref{eq:rho0AB}. Therefore, the last terms in Eq.~\eqref{eq:rho1example} and Eq.~\eqref{eq:rho1AA00}, respectively, do not contribute to the dynamics of \(\rho_s\).

For the two-photon coherences, we obtain (following a similar derivation as for the single-photon occupation): 
\begin{align}
    \partial_t\langle &20|\rho_s(t)|00 \rangle = -2\gamma_A\langle 20|\rho_s(t)|00 \rangle\nonumber\\
    &-\sqrt{8}V_{BA}^*\mathrm{e}^{i\omega_A \tau}\langle 10|\rho^{(1)L}_{s,0_B1_B}(t,t-\tau)|00 \rangle,
\end{align}
where we use \(0_B\) and \(1_B\) to indicate that the initial system-bath interaction involves the transition of cavity B from the one-photon state to the ground state. Analogously, \(\langle 02|\rho_s(t)|00 \rangle = (\langle 20|\rho_s(t)|00 \rangle)(A\leftrightarrow B)\) and
\begin{align}
    \partial_t\langle 11&|\rho_s(t)|00 \rangle = -(\gamma_A+\gamma_B)\langle 11|\rho_s(t)|00 \rangle\nonumber\\
    &-\sqrt{8}V_{BA}^*\mathrm{e}^{i\omega_A \tau}\langle 01|\rho^{(1)L}_{s,1_B2_B}(t,t-\tau)|00 \rangle\nonumber \\
    &-\sqrt{8}V_{AB}^*\mathrm{e}^{i\omega_B \tau}\langle 10|\rho^{(1)L}_{s,1_A2_A}(t,t-\tau)|00 \rangle\nonumber \\
    &-2 V_{BA}^*\mathrm{e}^{i\omega_A \tau}\langle 01|\rho^{(1)L}_{s,0_B1_B}(t,t-\tau)|00 \rangle\nonumber \\
    &-2 V_{AB}^*\mathrm{e}^{i\omega_B \tau}\langle 10|\rho^{(1)L}_{s,0_A1_A}(t,t-\tau)|00 \rangle.
\end{align}
The equations for the matrix elements of \(\rho^{(1)}\) in the rotating frame read (keeping only those terms that contribute to \(\rho_s\)):
\begin{align}
    &\partial_t\langle 10|\rho^{(1)L}_{s,0_B1_B}(t,t_1)|00 \rangle = \delta(t-t_1)\langle 11|\rho_s(t_1)|00 \rangle\nonumber\\
    &\qquad-\gamma_A\langle 10|\rho^{(1)L}_{s,0_B1_B}(t,t_1)|00 \rangle\nonumber\\
    &\qquad-\sqrt{8}V_{BA}^*\mathrm{e}^{i\omega_A\tau}\langle 01|\rho^{(1)L}_{s,1_B2_B}(t_1,t-\tau)|00 \rangle\nonumber\\
    &\qquad-2V_{BA}^*\mathrm{e}^{i\omega_A\tau}\langle 01|\rho^{(1)L}_{s,0_B1_B}(t_1,t-\tau)|00 \rangle,
\end{align}    
and
\begin{align}
    &\partial_t\langle 10|\rho^{(1)L}_{s,0_A1_A}(t,t_1)|00 \rangle = -\gamma_A\langle 10|\rho^{(1)L}_{s,0_A1_A}(t,t_1)|00 \rangle\nonumber\\
    &\qquad\qquad-2V_{BA}^*\mathrm{e}^{i\omega_A\tau}\langle 10|\rho^{(1)L}_{s,0_B1_B}(t_1,t-\tau)|00 \rangle,\nonumber\\
    &\partial_t\langle 10|\rho^{(1)L}_{s,1_A2_A}(t,t_1)|00 \rangle = \delta(t-t_1)\langle 20|\rho_s(t_1)|00 \rangle\nonumber\\
    &\qquad\qquad-\gamma_A\langle 10|\rho^{(1)L}_{s,1_A2_A}(t,t_1)|00 \rangle.
\end{align}
The remaining three matrix elements are again obtained by exchanging \(A\leftrightarrow B\).
The general equation for arbitrary states reads in the interaction picture:
\begin{align}
    &\partial_t \langle \nu_1|\rho_s(t)|\nu_2\rangle = -\sum_{\mu, \nu_3}\gamma_{\mu}\big[\langle \nu_1|\hat{a}_{\mu}^{\dagger}|\nu_3\rangle\langle \nu_3|\hat{a}_{\mu}|\nu_1\rangle\nonumber\\
    &\qquad\qquad\qquad\qquad+ \langle \nu_2|\hat{a}_{\mu}^{\dagger}|\nu_3\rangle\langle \nu_3|\hat{a}_{\mu}|\nu_2\rangle\big]\langle \nu_1|\rho_s(t)|\nu_2\rangle\nonumber \\
    &\quad+2\sum_{\mu,\nu_3\nu_4}\gamma_{\mu}\langle \nu_1|\hat{a}_{\mu}|\nu_3\rangle\langle \nu_3|\rho_s(t)|\nu_4\rangle\langle \nu_4|\hat{a}_{\mu}^{\dagger}|\nu_2\rangle\nonumber\\
    &\quad-2\sum_{\mu\eta,\nu_3\nu_4\nu_5}(1-\delta_{\mu\eta})V^*_{\eta\mu}\mathrm{e}^{i\omega_1\tau}\langle \nu_4|\hat{a}_{\eta}|\nu_5\rangle\nonumber\\
    &\qquad\qquad\times\big[\langle \nu_1|\hat{a}_{\mu}^{\dagger}|\nu_3\rangle\langle \nu_3|\rho^{(1),L}_{\nu_4\nu_5}(t,t-\tau)|\nu_2\rangle\nonumber\\
    &\qquad\qquad\quad-\langle \nu_1|\rho^{(1),L}_{\nu_4\nu_5}(t,t-\tau)|\nu_3\rangle\langle \nu_3|\hat{a}_{\mu}^{\dagger}|\nu_2\rangle\big]\nonumber\\
    &-2\sum_{\mu\eta,\nu_3\nu_4\nu_5}(1-\delta_{\mu\eta})V_{\eta\mu}\mathrm{e}^{-i\omega_1\tau}\langle\nu_4|\hat{a}^{\dagger}_{\eta}|\nu_5\rangle\nonumber\\
    &\qquad\qquad\times\big[\langle \nu_1|\rho^{(1),R}_{\nu_4\nu_5}(t,t-\tau)|\nu_3\rangle\langle \nu_3|\hat{a}_{\mu}|\nu_2\rangle\nonumber\\
    &\qquad\qquad\quad-\langle \nu_1|\hat{a}_{\mu}|\nu_3\rangle\langle\nu_3|\rho^{(1),R}_{\nu_4\nu_5}(t,t-\tau)|\nu_2\rangle\big].
\end{align}
Here, \(\hat{a}_{\mu}^{(\dagger)}\) are the QNM creation and annihilation operators from Eq.~\eqref{eq:qnmops}. \(\mu,\eta\) are system indices and \(\nu_i\) is an arbitrary system state. The general equation for \(\rho^{(1)}\) is obtained in the same manner and reads, in the rotating frame and keeping only terms that contribute to \(\rho_s\):
\begin{align}
    &\partial_t\langle\nu_1|\rho^{(1),L}_{\nu_3\nu_4}(t,t_1)|\nu_2\rangle = \delta(t-t_1)\delta_{\nu_1\nu_3}\langle \nu_4|\rho_s(t_1)|\nu_2\rangle\nonumber\\
    &-\sum_{\mu,\nu_5}\gamma_{\mu}\big[\langle \nu_1|\hat{a}_{\mu}^{\dagger}|\nu_5\rangle\langle \nu_5|\hat{a}_{\mu}|\nu_1\rangle
    + \langle \nu_2|\hat{a}_{\mu}^{\dagger}|\nu_5\rangle\langle \nu_5|\hat{a}_{\mu}|\nu_2\rangle\big]\nonumber\\
    &\qquad\qquad\qquad\qquad\qquad\qquad\qquad\times\langle\nu_1|\rho^{(1),L}_{\nu_3\nu_4}(t,t_1)|\nu_2\rangle\nonumber \\
    &+2\sum_{\mu,\nu_5\nu_6}\gamma_{\mu}\langle \nu_1|\hat{a}_{\mu}|\nu_5\rangle\langle\nu_5|\rho^{(1),L}_{\nu_3\nu_4}(t,t_1)|\nu_6\rangle\langle \nu_6|\hat{a}_{\mu}^{\dagger}|\nu_2\rangle\nonumber\\
    &-2\sum_{\mu\eta,\nu_5\nu_6\nu_7}(1-\delta_{\mu\eta})V^*_{\eta\mu}\mathrm{e}^{i\omega_1\tau}\langle \nu_6|\hat{a}_{\eta}|\nu_7\rangle\nonumber\\
    &\qquad\times\big[\delta_{\nu_3\nu_5}\langle \nu_1|\hat{a}_{\mu}^{\dagger}|\nu_3\rangle\langle \nu_4|\rho^{(1),L}_{\nu_6\nu_7}(t_1,t-\tau)|\nu_2\rangle\nonumber\\
    &\qquad\quad-\delta_{\nu_1\nu_3}\langle \nu_4|\rho^{(1),L}_{\nu_6\nu_7}(t_1,t-\tau)|\nu_5\rangle\langle \nu_5|\hat{a}_{\mu}^{\dagger}|\nu_2\rangle\big]\nonumber\\
    &-2\sum_{\mu\eta,\nu_5\nu_6\nu_7}(1-\delta_{\mu\eta})V_{\eta\mu}\mathrm{e}^{-i\omega_1\tau}\langle\nu_6|\hat{a}^{\dagger}_{\eta}|\nu_7\rangle\nonumber\\
    &\qquad\times\big[\delta_{\nu_1\nu_3}\langle \nu_4|\rho^{(1),R}_{\nu_6\nu_7}(t_1,t-\tau)|\nu_3\rangle\langle \nu_3|\hat{a}_{\mu}|\nu_2\rangle\nonumber\\
    &\qquad\quad-\delta_{\nu_3\nu_5}\langle \nu_1|\hat{a}_{\mu}|\nu_3\rangle\langle\nu_4|\rho^{(1),R}_{\nu_6\nu_7}(t_1,t-\tau)|\nu_2\rangle\big].
\end{align}

\section{Wave function approach}
\label{app3}
For initially one excitation in slab A from Fig.~\ref{fig:model}(b), the general wave function has the form
\begin{align} \label{eq:psistate}
    |\psi\rangle = N_A|A\rangle|0\rangle + N_B|B\rangle|0\rangle +\int\mathrm{d}x\int_0^{\infty}\mathrm{d}\omega N_{x,\omega}|0\rangle|x,\omega\rangle.
\end{align}
The first state in the product state refers to the system, and the second is the bath with continuous spatial and frequency indices \(x,\omega\). \(N\) is the time-dependent amplitude of a particular state, with the initial conditions \(N_A(0) = 1,\, N_B(0) =  N_{x,\omega}(0) = 0\). In the interaction picture, the dynamics of the states are governed by the Schrödinger equation with the system-bath interaction Hamiltonian from Eq.~\eqref{eq:QNMHamiltonian}. The QNM and bath operators carry the free evolution of the system and bath: \(\hat{a}_{\mu}(t) = \mathrm{e}^{-i\omega_{\mu}t}\hat{a}_{\mu}\) and
\(\hat{c}(x,\omega,t) = \mathrm{e}^{-i\omega t}\hat{c}(x,\omega)\).\\
Multiplying the Schrödinger equation for \eqref{eq:psistate} with \(\langle 0|\langle A|\) from the left yields an equation for \(N_A\):
\begin{align} \label{eq:NAeq}
    i\hbar \partial_t N_A(t) = \hbar\int\mathrm{d}x\int_0^{\infty}\mathrm{d}\omega N_{x,\omega}g_A(x,\omega)\mathrm{e}^{-i\omega t}\mathrm{e}^{i\omega_A t}.
\end{align}
Similarly, we obtain the equation for \(N_{x,\omega}\):
\begin{align*}
    i\hbar \partial_t N_{x,\omega}(t) = \hbar &\left(N_{A}g^*_A(x,\omega)\mathrm{e}^{-i\omega_A t}\right.\\
    &\left.+N_Bg^*_B(x,\omega)\mathrm{e}^{-i\omega_B t} \right)\mathrm{e}^{i\omega t},
\end{align*}
which we integrate formally and insert the result back into Eq.~\eqref{eq:NAeq} to find:
\begin{align}
    \partial_t N_A(t) = -\frac{1}{\hbar^2}&\int_{0}^{t}\mathrm{d}t' \Big(C_{AA}(t-t')N_A(t')\nonumber\\ 
    &+\mathrm{e}^{i\omega_A t-i\omega_B t'}C_{AB}(t-t')N_B(t')\Big),
\end{align}
where we have inserted the definition of the QNM correlation function from Eq.~\eqref{eq:corrfuncgeneral}. Using Eq.~\eqref{eq:corrfunc} and \(\omega_A=\omega_B\), we arrive at:
\begin{align}
    \partial_t N_A(t) = -\gamma_A N_A(t)-2V_{BA}^*\mathrm{e}^{i\omega_B\tau}N_B(t-\tau)\Theta(t-\tau).
\end{align}
An analogous derivation for \(N_B\) yields a similar equation, with the indices switched \((A\leftrightarrow B)\). The density matrix elements are calculated by multiplying the amplitudes with their complex conjugates, e.g., \(\langle A|\rho_s|A\rangle = |N_A|^2\).
\bibliography{heombiblio}

\begin{thebibliography}{60}%
\makeatletter
\providecommand \@ifxundefined [1]{%
 \@ifx{#1\undefined}
}%
\providecommand \@ifnum [1]{%
 \ifnum #1\expandafter \@firstoftwo
 \else \expandafter \@secondoftwo
 \fi
}%
\providecommand \@ifx [1]{%
 \ifx #1\expandafter \@firstoftwo
 \else \expandafter \@secondoftwo
 \fi
}%
\providecommand \natexlab [1]{#1}%
\providecommand \enquote  [1]{``#1''}%
\providecommand \bibnamefont  [1]{#1}%
\providecommand \bibfnamefont [1]{#1}%
\providecommand \citenamefont [1]{#1}%
\providecommand \href@noop [0]{\@secondoftwo}%
\providecommand \href [0]{\begingroup \@sanitize@url \@href}%
\providecommand \@href[1]{\@@startlink{#1}\@@href}%
\providecommand \@@href[1]{\endgroup#1\@@endlink}%
\providecommand \@sanitize@url [0]{\catcode `\\12\catcode `\$12\catcode
  `\&12\catcode `\#12\catcode `\^12\catcode `\_12\catcode `\%12\relax}%
\providecommand \@@startlink[1]{}%
\providecommand \@@endlink[0]{}%
\providecommand \url  [0]{\begingroup\@sanitize@url \@url }%
\providecommand \@url [1]{\endgroup\@href {#1}{\urlprefix }}%
\providecommand \urlprefix  [0]{URL }%
\providecommand \Eprint [0]{\href }%
\providecommand \doibase [0]{https://doi.org/}%
\providecommand \selectlanguage [0]{\@gobble}%
\providecommand \bibinfo  [0]{\@secondoftwo}%
\providecommand \bibfield  [0]{\@secondoftwo}%
\providecommand \translation [1]{[#1]}%
\providecommand \BibitemOpen [0]{}%
\providecommand \bibitemStop [0]{}%
\providecommand \bibitemNoStop [0]{.\EOS\space}%
\providecommand \EOS [0]{\spacefactor3000\relax}%
\providecommand \BibitemShut  [1]{\csname bibitem#1\endcsname}%
\let\auto@bib@innerbib\@empty
\bibitem [{\citenamefont {Tanimura}\ and\ \citenamefont
  {Kubo}(1989)}]{originalheom}%
  \BibitemOpen
  \bibfield  {author} {\bibinfo {author} {\bibfnamefont {Y.}~\bibnamefont
  {Tanimura}}\ and\ \bibinfo {author} {\bibfnamefont {R.}~\bibnamefont
  {Kubo}},\ }\bibfield  {title} {\bibinfo {title} {Time evolution of a quantum
  system in contact with a nearly gaussian-markoffian noise bath},\ }\href
  {https://doi.org/10.1143/JPSJ.58.101} {\bibfield  {journal} {\bibinfo
  {journal} {Journal of the Physical Society of Japan}\ }\textbf {\bibinfo
  {volume} {58}},\ \bibinfo {pages} {101} (\bibinfo {year} {1989})}\BibitemShut
  {NoStop}%
\bibitem [{\citenamefont {Tanimura}(2020)}]{tanimura2020numerically}%
  \BibitemOpen
  \bibfield  {author} {\bibinfo {author} {\bibfnamefont {Y.}~\bibnamefont
  {Tanimura}},\ }\bibfield  {title} {\bibinfo {title} {Numerically “exact”
  approach to open quantum dynamics: The hierarchical equations of motion
  (heom)},\ }\href@noop {} {\bibfield  {journal} {\bibinfo  {journal} {The
  Journal of chemical physics}\ }\textbf {\bibinfo {volume} {153}},\ \bibinfo
  {pages} {020901} (\bibinfo {year} {2020})}\BibitemShut {NoStop}%
\bibitem [{\citenamefont {Ye}\ \emph {et~al.}(2016)\citenamefont {Ye},
  \citenamefont {Wang}, \citenamefont {Hou}, \citenamefont {Xu}, \citenamefont
  {Zheng},\ and\ \citenamefont {Yan}}]{heomquick}%
  \BibitemOpen
  \bibfield  {author} {\bibinfo {author} {\bibfnamefont {L.}~\bibnamefont
  {Ye}}, \bibinfo {author} {\bibfnamefont {X.}~\bibnamefont {Wang}}, \bibinfo
  {author} {\bibfnamefont {D.}~\bibnamefont {Hou}}, \bibinfo {author}
  {\bibfnamefont {R.-X.}\ \bibnamefont {Xu}}, \bibinfo {author} {\bibfnamefont
  {X.}~\bibnamefont {Zheng}},\ and\ \bibinfo {author} {\bibfnamefont
  {Y.}~\bibnamefont {Yan}},\ }\bibfield  {title} {\bibinfo {title} {Heom-quick:
  a program for accurate, efficient, and universal characterization of strongly
  correlated quantum impurity systems},\ }\href
  {https://doi.org/https://doi.org/10.1002/wcms.1269} {\bibfield  {journal}
  {\bibinfo  {journal} {WIREs Computational Molecular Science}\ }\textbf
  {\bibinfo {volume} {6}},\ \bibinfo {pages} {608} (\bibinfo {year}
  {2016})}\BibitemShut {NoStop}%
\bibitem [{\citenamefont {Lambert}\ \emph {et~al.}(2020)\citenamefont
  {Lambert}, \citenamefont {Raheja}, \citenamefont {Ahmed}, \citenamefont
  {Pitchford},\ and\ \citenamefont {Nori}}]{lambert2020bofin}%
  \BibitemOpen
  \bibfield  {author} {\bibinfo {author} {\bibfnamefont {N.}~\bibnamefont
  {Lambert}}, \bibinfo {author} {\bibfnamefont {T.}~\bibnamefont {Raheja}},
  \bibinfo {author} {\bibfnamefont {S.}~\bibnamefont {Ahmed}}, \bibinfo
  {author} {\bibfnamefont {A.}~\bibnamefont {Pitchford}},\ and\ \bibinfo
  {author} {\bibfnamefont {F.}~\bibnamefont {Nori}},\ }\bibfield  {title}
  {\bibinfo {title} {Bofin-heom: A bosonic and fermionic numerical
  hierarchical-equations-of-motion library with applications in
  light-harvesting, quantum control, and single-molecule electronics},\
  }\href@noop {} {\bibfield  {journal} {\bibinfo  {journal} {arXiv preprint
  arXiv:2010.10806}\ } (\bibinfo {year} {2020})}\BibitemShut {NoStop}%
\bibitem [{\citenamefont {Kramer}\ \emph
  {et~al.}(2018{\natexlab{a}})\citenamefont {Kramer}, \citenamefont {Noack},
  \citenamefont {Reinefeld}, \citenamefont {Rodr{\'\i}guez},\ and\
  \citenamefont {Zelinskyy}}]{kramer2018efficient}%
  \BibitemOpen
  \bibfield  {author} {\bibinfo {author} {\bibfnamefont {T.}~\bibnamefont
  {Kramer}}, \bibinfo {author} {\bibfnamefont {M.}~\bibnamefont {Noack}},
  \bibinfo {author} {\bibfnamefont {A.}~\bibnamefont {Reinefeld}}, \bibinfo
  {author} {\bibfnamefont {M.}~\bibnamefont {Rodr{\'\i}guez}},\ and\ \bibinfo
  {author} {\bibfnamefont {Y.}~\bibnamefont {Zelinskyy}},\ }\bibfield  {title}
  {\bibinfo {title} {Efficient calculation of open quantum system dynamics and
  time-resolved spectroscopy with distributed memory heom (dm-heom)},\
  }\href@noop {} {\bibfield  {journal} {\bibinfo  {journal} {Journal of
  Computational Chemistry}\ }\textbf {\bibinfo {volume} {39}},\ \bibinfo
  {pages} {1779} (\bibinfo {year} {2018}{\natexlab{a}})}\BibitemShut {NoStop}%
\bibitem [{\citenamefont {Seibt}\ and\ \citenamefont
  {Kühn}(2021)}]{seibt2021strong}%
  \BibitemOpen
  \bibfield  {author} {\bibinfo {author} {\bibfnamefont {J.}~\bibnamefont
  {Seibt}}\ and\ \bibinfo {author} {\bibfnamefont {O.}~\bibnamefont {Kühn}},\
  }\bibfield  {title} {\bibinfo {title} {Strong exciton-vibrational coupling in
  molecular assemblies. dynamics using the polaron transformation in heom
  space},\ }\href@noop {} {\bibfield  {journal} {\bibinfo  {journal} {The
  Journal of Physical Chemistry A}\ }\textbf {\bibinfo {volume} {125}},\
  \bibinfo {pages} {7052} (\bibinfo {year} {2021})}\BibitemShut {NoStop}%
\bibitem [{\citenamefont {Kramer}\ \emph
  {et~al.}(2018{\natexlab{b}})\citenamefont {Kramer}, \citenamefont {Noack},
  \citenamefont {Reimers}, \citenamefont {Reinefeld}, \citenamefont
  {Rodr{\'\i}guez},\ and\ \citenamefont {Yin}}]{kramer2018energy}%
  \BibitemOpen
  \bibfield  {author} {\bibinfo {author} {\bibfnamefont {T.}~\bibnamefont
  {Kramer}}, \bibinfo {author} {\bibfnamefont {M.}~\bibnamefont {Noack}},
  \bibinfo {author} {\bibfnamefont {J.~R.}\ \bibnamefont {Reimers}}, \bibinfo
  {author} {\bibfnamefont {A.}~\bibnamefont {Reinefeld}}, \bibinfo {author}
  {\bibfnamefont {M.}~\bibnamefont {Rodr{\'\i}guez}},\ and\ \bibinfo {author}
  {\bibfnamefont {S.}~\bibnamefont {Yin}},\ }\bibfield  {title} {\bibinfo
  {title} {Energy flow in the photosystem i supercomplex: Comparison of
  approximative theories with dm-heom},\ }\href@noop {} {\bibfield  {journal}
  {\bibinfo  {journal} {Chemical Physics}\ }\textbf {\bibinfo {volume} {515}},\
  \bibinfo {pages} {262} (\bibinfo {year} {2018}{\natexlab{b}})}\BibitemShut
  {NoStop}%
\bibitem [{\citenamefont {Oulton}\ \emph {et~al.}(2008)\citenamefont {Oulton},
  \citenamefont {Sorger}, \citenamefont {Genov}, \citenamefont {Pile},\ and\
  \citenamefont {Zhang}}]{oulton2008hybrid}%
  \BibitemOpen
  \bibfield  {author} {\bibinfo {author} {\bibfnamefont {R.~F.}\ \bibnamefont
  {Oulton}}, \bibinfo {author} {\bibfnamefont {V.~J.}\ \bibnamefont {Sorger}},
  \bibinfo {author} {\bibfnamefont {D.}~\bibnamefont {Genov}}, \bibinfo
  {author} {\bibfnamefont {D.}~\bibnamefont {Pile}},\ and\ \bibinfo {author}
  {\bibfnamefont {X.}~\bibnamefont {Zhang}},\ }\bibfield  {title} {\bibinfo
  {title} {A hybrid plasmonic waveguide for subwavelength confinement and
  long-range propagation},\ }\href@noop {} {\bibfield  {journal} {\bibinfo
  {journal} {Nature Photonics}\ }\textbf {\bibinfo {volume} {2}},\ \bibinfo
  {pages} {496} (\bibinfo {year} {2008})}\BibitemShut {NoStop}%
\bibitem [{\citenamefont {Stockman}(2004)}]{stockman2004nanofocusing}%
  \BibitemOpen
  \bibfield  {author} {\bibinfo {author} {\bibfnamefont {M.~I.}\ \bibnamefont
  {Stockman}},\ }\bibfield  {title} {\bibinfo {title} {Nanofocusing of optical
  energy in tapered plasmonic waveguides},\ }\href@noop {} {\bibfield
  {journal} {\bibinfo  {journal} {Physical review letters}\ }\textbf {\bibinfo
  {volume} {93}},\ \bibinfo {pages} {137404} (\bibinfo {year}
  {2004})}\BibitemShut {NoStop}%
\bibitem [{\citenamefont {Orieux}\ \emph {et~al.}(2017)\citenamefont {Orieux},
  \citenamefont {Versteegh}, \citenamefont {J{\"o}ns},\ and\ \citenamefont
  {Ducci}}]{orieux2017semiconductor}%
  \BibitemOpen
  \bibfield  {author} {\bibinfo {author} {\bibfnamefont {A.}~\bibnamefont
  {Orieux}}, \bibinfo {author} {\bibfnamefont {M.~A.}\ \bibnamefont
  {Versteegh}}, \bibinfo {author} {\bibfnamefont {K.~D.}\ \bibnamefont
  {J{\"o}ns}},\ and\ \bibinfo {author} {\bibfnamefont {S.}~\bibnamefont
  {Ducci}},\ }\bibfield  {title} {\bibinfo {title} {Semiconductor devices for
  entangled photon pair generation: a review},\ }\href@noop {} {\bibfield
  {journal} {\bibinfo  {journal} {Reports on Progress in Physics}\ }\textbf
  {\bibinfo {volume} {80}},\ \bibinfo {pages} {076001} (\bibinfo {year}
  {2017})}\BibitemShut {NoStop}%
\bibitem [{\citenamefont {Wei{\ss}}\ and\ \citenamefont
  {Krenner}(2018)}]{weiss2018interfacing}%
  \BibitemOpen
  \bibfield  {author} {\bibinfo {author} {\bibfnamefont {M.}~\bibnamefont
  {Wei{\ss}}}\ and\ \bibinfo {author} {\bibfnamefont {H.~J.}\ \bibnamefont
  {Krenner}},\ }\bibfield  {title} {\bibinfo {title} {Interfacing quantum
  emitters with propagating surface acoustic waves},\ }\href@noop {} {\bibfield
   {journal} {\bibinfo  {journal} {Journal of Physics D: Applied Physics}\
  }\textbf {\bibinfo {volume} {51}},\ \bibinfo {pages} {373001} (\bibinfo
  {year} {2018})}\BibitemShut {NoStop}%
\bibitem [{\citenamefont {Jayakumar}\ \emph {et~al.}(2014)\citenamefont
  {Jayakumar}, \citenamefont {Predojevi{\'c}}, \citenamefont {Kauten},
  \citenamefont {Huber}, \citenamefont {Solomon},\ and\ \citenamefont
  {Weihs}}]{jayakumar2014time}%
  \BibitemOpen
  \bibfield  {author} {\bibinfo {author} {\bibfnamefont {H.}~\bibnamefont
  {Jayakumar}}, \bibinfo {author} {\bibfnamefont {A.}~\bibnamefont
  {Predojevi{\'c}}}, \bibinfo {author} {\bibfnamefont {T.}~\bibnamefont
  {Kauten}}, \bibinfo {author} {\bibfnamefont {T.}~\bibnamefont {Huber}},
  \bibinfo {author} {\bibfnamefont {G.~S.}\ \bibnamefont {Solomon}},\ and\
  \bibinfo {author} {\bibfnamefont {G.}~\bibnamefont {Weihs}},\ }\bibfield
  {title} {\bibinfo {title} {Time-bin entangled photons from a quantum dot},\
  }\href@noop {} {\bibfield  {journal} {\bibinfo  {journal} {Nature
  communications}\ }\textbf {\bibinfo {volume} {5}},\ \bibinfo {pages} {1}
  (\bibinfo {year} {2014})}\BibitemShut {NoStop}%
\bibitem [{\citenamefont {Carmele}\ and\ \citenamefont
  {Reitzenstein}(2019)}]{CarmeleReitzenstein}%
  \BibitemOpen
  \bibfield  {author} {\bibinfo {author} {\bibfnamefont {A.}~\bibnamefont
  {Carmele}}\ and\ \bibinfo {author} {\bibfnamefont {S.}~\bibnamefont
  {Reitzenstein}},\ }\bibfield  {title} {\bibinfo {title} {Non-markovian
  features in semiconductor quantum optics: quantifying the role of phonons in
  experiment and theory},\ }\href
  {https://doi.org/doi:10.1515/nanoph-2018-0222} {\bibfield  {journal}
  {\bibinfo  {journal} {Nanophotonics}\ }\textbf {\bibinfo {volume} {8}},\
  \bibinfo {pages} {655} (\bibinfo {year} {2019})}\BibitemShut {NoStop}%
\bibitem [{\citenamefont {Pichler}\ and\ \citenamefont
  {Zoller}(2016)}]{PhysRevLett.116.093601}%
  \BibitemOpen
  \bibfield  {author} {\bibinfo {author} {\bibfnamefont {H.}~\bibnamefont
  {Pichler}}\ and\ \bibinfo {author} {\bibfnamefont {P.}~\bibnamefont
  {Zoller}},\ }\bibfield  {title} {\bibinfo {title} {Photonic circuits with
  time delays and quantum feedback},\ }\href
  {https://doi.org/10.1103/PhysRevLett.116.093601} {\bibfield  {journal}
  {\bibinfo  {journal} {Phys. Rev. Lett.}\ }\textbf {\bibinfo {volume} {116}},\
  \bibinfo {pages} {093601} (\bibinfo {year} {2016})}\BibitemShut {NoStop}%
\bibitem [{\citenamefont {Kaestle}\ \emph {et~al.}(2021)\citenamefont
  {Kaestle}, \citenamefont {Finsterhoelzl}, \citenamefont {Knorr},\ and\
  \citenamefont {Carmele}}]{kaestle2020protected}%
  \BibitemOpen
  \bibfield  {author} {\bibinfo {author} {\bibfnamefont {O.}~\bibnamefont
  {Kaestle}}, \bibinfo {author} {\bibfnamefont {R.}~\bibnamefont
  {Finsterhoelzl}}, \bibinfo {author} {\bibfnamefont {A.}~\bibnamefont
  {Knorr}},\ and\ \bibinfo {author} {\bibfnamefont {A.}~\bibnamefont
  {Carmele}},\ }\bibfield  {title} {\bibinfo {title} {Continuous and
  time-discrete non-markovian system-reservoir interactions: Dissipative
  coherent quantum feedback in liouville space},\ }\href
  {https://doi.org/10.1103/PhysRevResearch.3.023168} {\bibfield  {journal}
  {\bibinfo  {journal} {Phys. Rev. Research}\ }\textbf {\bibinfo {volume}
  {3}},\ \bibinfo {pages} {023168} (\bibinfo {year} {2021})}\BibitemShut
  {NoStop}%
\bibitem [{\citenamefont {Arranz~Regidor}\ \emph {et~al.}(2021)\citenamefont
  {Arranz~Regidor}, \citenamefont {Crowder}, \citenamefont {Carmichael},\ and\
  \citenamefont {Hughes}}]{PhysRevResearch.3.023030}%
  \BibitemOpen
  \bibfield  {author} {\bibinfo {author} {\bibfnamefont {S.}~\bibnamefont
  {Arranz~Regidor}}, \bibinfo {author} {\bibfnamefont {G.}~\bibnamefont
  {Crowder}}, \bibinfo {author} {\bibfnamefont {H.}~\bibnamefont
  {Carmichael}},\ and\ \bibinfo {author} {\bibfnamefont {S.}~\bibnamefont
  {Hughes}},\ }\bibfield  {title} {\bibinfo {title} {Modeling quantum
  light-matter interactions in waveguide qed with retardation, nonlinear
  interactions, and a time-delayed feedback: Matrix product states versus a
  space-discretized waveguide model},\ }\href
  {https://doi.org/10.1103/PhysRevResearch.3.023030} {\bibfield  {journal}
  {\bibinfo  {journal} {Phys. Rev. Research}\ }\textbf {\bibinfo {volume}
  {3}},\ \bibinfo {pages} {023030} (\bibinfo {year} {2021})}\BibitemShut
  {NoStop}%
\bibitem [{\citenamefont {Richter}\ and\ \citenamefont
  {Hughes}(2022)}]{PhysRevLett.128.167403}%
  \BibitemOpen
  \bibfield  {author} {\bibinfo {author} {\bibfnamefont {M.}~\bibnamefont
  {Richter}}\ and\ \bibinfo {author} {\bibfnamefont {S.}~\bibnamefont
  {Hughes}},\ }\bibfield  {title} {\bibinfo {title} {Enhanced tempo algorithm
  for quantum path integrals with off-diagonal system-bath coupling:
  Applications to photonic quantum networks},\ }\href
  {https://doi.org/10.1103/PhysRevLett.128.167403} {\bibfield  {journal}
  {\bibinfo  {journal} {Phys. Rev. Lett.}\ }\textbf {\bibinfo {volume} {128}},\
  \bibinfo {pages} {167403} (\bibinfo {year} {2022})}\BibitemShut {NoStop}%
\bibitem [{\citenamefont {Caldeira}\ and\ \citenamefont
  {Leggett}(1983)}]{CALDEIRA1983587}%
  \BibitemOpen
  \bibfield  {author} {\bibinfo {author} {\bibfnamefont {A.}~\bibnamefont
  {Caldeira}}\ and\ \bibinfo {author} {\bibfnamefont {A.}~\bibnamefont
  {Leggett}},\ }\bibfield  {title} {\bibinfo {title} {Path integral approach to
  quantum brownian motion},\ }\href
  {https://doi.org/https://doi.org/10.1016/0378-4371(83)90013-4} {\bibfield
  {journal} {\bibinfo  {journal} {Physica A}\ }\textbf {\bibinfo {volume}
  {121}},\ \bibinfo {pages} {587} (\bibinfo {year} {1983})}\BibitemShut
  {NoStop}%
\bibitem [{\citenamefont {Tanimura}\ and\ \citenamefont
  {Mukamel}(1993)}]{tanimura1993real}%
  \BibitemOpen
  \bibfield  {author} {\bibinfo {author} {\bibfnamefont {Y.}~\bibnamefont
  {Tanimura}}\ and\ \bibinfo {author} {\bibfnamefont {S.}~\bibnamefont
  {Mukamel}},\ }\bibfield  {title} {\bibinfo {title} {Real-time path-integral
  approach to quantum coherence and dephasing in nonadiabatic transitions and
  nonlinear optical response},\ }\href@noop {} {\bibfield  {journal} {\bibinfo
  {journal} {Phys. Rev. E}\ }\textbf {\bibinfo {volume} {47}},\ \bibinfo
  {pages} {118} (\bibinfo {year} {1993})}\BibitemShut {NoStop}%
\bibitem [{\citenamefont {Makri}\ and\ \citenamefont
  {Makarov}(1995{\natexlab{a}})}]{makri1995tensor}%
  \BibitemOpen
  \bibfield  {author} {\bibinfo {author} {\bibfnamefont {N.}~\bibnamefont
  {Makri}}\ and\ \bibinfo {author} {\bibfnamefont {D.~E.}\ \bibnamefont
  {Makarov}},\ }\bibfield  {title} {\bibinfo {title} {Tensor propagator for
  iterative quantum time evolution of reduced density matrices. i. theory},\
  }\href@noop {} {\bibfield  {journal} {\bibinfo  {journal} {J. Chem. Phys.}\
  }\textbf {\bibinfo {volume} {102}},\ \bibinfo {pages} {4600} (\bibinfo {year}
  {1995}{\natexlab{a}})}\BibitemShut {NoStop}%
\bibitem [{\citenamefont {Makri}\ and\ \citenamefont
  {Makarov}(1995{\natexlab{b}})}]{makri1995tensor2}%
  \BibitemOpen
  \bibfield  {author} {\bibinfo {author} {\bibfnamefont {N.}~\bibnamefont
  {Makri}}\ and\ \bibinfo {author} {\bibfnamefont {D.~E.}\ \bibnamefont
  {Makarov}},\ }\bibfield  {title} {\bibinfo {title} {Tensor propagator for
  iterative quantum time evolution of reduced density matrices. ii. numerical
  methodology},\ }\href@noop {} {\bibfield  {journal} {\bibinfo  {journal} {J.
  Chem. Phys.}\ }\textbf {\bibinfo {volume} {102}},\ \bibinfo {pages} {4611}
  (\bibinfo {year} {1995}{\natexlab{b}})}\BibitemShut {NoStop}%
\bibitem [{\citenamefont {Vagov}\ \emph {et~al.}(2011)\citenamefont {Vagov},
  \citenamefont {Croitoru}, \citenamefont {Gl{\"a}ssl}, \citenamefont {Axt},\
  and\ \citenamefont {Kuhn}}]{vagov2011real}%
  \BibitemOpen
  \bibfield  {author} {\bibinfo {author} {\bibfnamefont {A.}~\bibnamefont
  {Vagov}}, \bibinfo {author} {\bibfnamefont {M.~D.}\ \bibnamefont {Croitoru}},
  \bibinfo {author} {\bibfnamefont {M.}~\bibnamefont {Gl{\"a}ssl}}, \bibinfo
  {author} {\bibfnamefont {V.~M.}\ \bibnamefont {Axt}},\ and\ \bibinfo {author}
  {\bibfnamefont {T.}~\bibnamefont {Kuhn}},\ }\bibfield  {title} {\bibinfo
  {title} {Real-time path integrals for quantum dots: Quantum dissipative
  dynamics with superohmic environment coupling},\ }\href@noop {} {\bibfield
  {journal} {\bibinfo  {journal} {Phys. Rev. B}\ }\textbf {\bibinfo {volume}
  {83}},\ \bibinfo {pages} {094303} (\bibinfo {year} {2011})}\BibitemShut
  {NoStop}%
\bibitem [{\citenamefont {Strathearn}\ \emph {et~al.}(2017)\citenamefont
  {Strathearn}, \citenamefont {Lovett},\ and\ \citenamefont
  {Kirton}}]{strathearn2017efficient}%
  \BibitemOpen
  \bibfield  {author} {\bibinfo {author} {\bibfnamefont {A.}~\bibnamefont
  {Strathearn}}, \bibinfo {author} {\bibfnamefont {B.~W.}\ \bibnamefont
  {Lovett}},\ and\ \bibinfo {author} {\bibfnamefont {P.}~\bibnamefont
  {Kirton}},\ }\bibfield  {title} {\bibinfo {title} {Efficient real-time path
  integrals for non-markovian spin-boson models},\ }\href@noop {} {\bibfield
  {journal} {\bibinfo  {journal} {New Journal of Physics}\ }\textbf {\bibinfo
  {volume} {19}},\ \bibinfo {pages} {093009} (\bibinfo {year}
  {2017})}\BibitemShut {NoStop}%
\bibitem [{\citenamefont {Strathearn}\ \emph {et~al.}(2018)\citenamefont
  {Strathearn}, \citenamefont {Kirton}, \citenamefont {Kilda}, \citenamefont
  {Keeling},\ and\ \citenamefont {Lovett}}]{strathearn2018TEMPO}%
  \BibitemOpen
  \bibfield  {author} {\bibinfo {author} {\bibfnamefont {A.}~\bibnamefont
  {Strathearn}}, \bibinfo {author} {\bibfnamefont {P.}~\bibnamefont {Kirton}},
  \bibinfo {author} {\bibfnamefont {D.}~\bibnamefont {Kilda}}, \bibinfo
  {author} {\bibfnamefont {J.}~\bibnamefont {Keeling}},\ and\ \bibinfo {author}
  {\bibfnamefont {B.~W.}\ \bibnamefont {Lovett}},\ }\bibfield  {title}
  {\bibinfo {title} {Efficient non-markovian quantum dynamics using
  time-evolving matrix product operators},\ }\href@noop {} {\bibfield
  {journal} {\bibinfo  {journal} {Nature communications}\ }\textbf {\bibinfo
  {volume} {9}},\ \bibinfo {pages} {3322} (\bibinfo {year} {2018})}\BibitemShut
  {NoStop}%
\bibitem [{\citenamefont {Gribben}\ \emph {et~al.}(2022)\citenamefont
  {Gribben}, \citenamefont {Rouse}, \citenamefont {Iles-Smith}, \citenamefont
  {Strathearn}, \citenamefont {Maguire}, \citenamefont {Kirton}, \citenamefont
  {Nazir}, \citenamefont {Gauger},\ and\ \citenamefont
  {Lovett}}]{gribben2021exact}%
  \BibitemOpen
  \bibfield  {author} {\bibinfo {author} {\bibfnamefont {D.}~\bibnamefont
  {Gribben}}, \bibinfo {author} {\bibfnamefont {D.~M.}\ \bibnamefont {Rouse}},
  \bibinfo {author} {\bibfnamefont {J.}~\bibnamefont {Iles-Smith}}, \bibinfo
  {author} {\bibfnamefont {A.}~\bibnamefont {Strathearn}}, \bibinfo {author}
  {\bibfnamefont {H.}~\bibnamefont {Maguire}}, \bibinfo {author} {\bibfnamefont
  {P.}~\bibnamefont {Kirton}}, \bibinfo {author} {\bibfnamefont
  {A.}~\bibnamefont {Nazir}}, \bibinfo {author} {\bibfnamefont {E.~M.}\
  \bibnamefont {Gauger}},\ and\ \bibinfo {author} {\bibfnamefont {B.~W.}\
  \bibnamefont {Lovett}},\ }\bibfield  {title} {\bibinfo {title} {Exact
  dynamics of nonadditive environments in non-markovian open quantum systems},\
  }\href {https://doi.org/10.1103/PRXQuantum.3.010321} {\bibfield  {journal}
  {\bibinfo  {journal} {PRX Quantum}\ }\textbf {\bibinfo {volume} {3}},\
  \bibinfo {pages} {010321} (\bibinfo {year} {2022})}\BibitemShut {NoStop}%
\bibitem [{\citenamefont {Cygorek}\ \emph {et~al.}(2022)\citenamefont
  {Cygorek}, \citenamefont {Cosacchi}, \citenamefont {Vagov}, \citenamefont
  {Axt}, \citenamefont {Lovett}, \citenamefont {Keeling},\ and\ \citenamefont
  {Gauger}}]{cygorek2021numerically}%
  \BibitemOpen
  \bibfield  {author} {\bibinfo {author} {\bibfnamefont {M.}~\bibnamefont
  {Cygorek}}, \bibinfo {author} {\bibfnamefont {M.}~\bibnamefont {Cosacchi}},
  \bibinfo {author} {\bibfnamefont {A.}~\bibnamefont {Vagov}}, \bibinfo
  {author} {\bibfnamefont {V.~M.}\ \bibnamefont {Axt}}, \bibinfo {author}
  {\bibfnamefont {B.~W.}\ \bibnamefont {Lovett}}, \bibinfo {author}
  {\bibfnamefont {J.}~\bibnamefont {Keeling}},\ and\ \bibinfo {author}
  {\bibfnamefont {E.~M.}\ \bibnamefont {Gauger}},\ }\bibfield  {title}
  {\bibinfo {title} {Simulation of open quantum systems by automated
  compression of arbitrary environments},\ }\href@noop {} {\bibfield  {journal}
  {\bibinfo  {journal} {Nature Physics}\ ,\ \bibinfo {pages} {1}} (\bibinfo
  {year} {2022})}\BibitemShut {NoStop}%
\bibitem [{\citenamefont {Prior}\ \emph {et~al.}(2013)\citenamefont {Prior},
  \citenamefont {de~Vega}, \citenamefont {Chin}, \citenamefont {Huelga},\ and\
  \citenamefont {Plenio}}]{PhysRevA.87.013428}%
  \BibitemOpen
  \bibfield  {author} {\bibinfo {author} {\bibfnamefont {J.}~\bibnamefont
  {Prior}}, \bibinfo {author} {\bibfnamefont {I.}~\bibnamefont {de~Vega}},
  \bibinfo {author} {\bibfnamefont {A.~W.}\ \bibnamefont {Chin}}, \bibinfo
  {author} {\bibfnamefont {S.~F.}\ \bibnamefont {Huelga}},\ and\ \bibinfo
  {author} {\bibfnamefont {M.~B.}\ \bibnamefont {Plenio}},\ }\bibfield  {title}
  {\bibinfo {title} {Quantum dynamics in photonic crystals},\ }\href
  {https://doi.org/10.1103/PhysRevA.87.013428} {\bibfield  {journal} {\bibinfo
  {journal} {Phys. Rev. A}\ }\textbf {\bibinfo {volume} {87}},\ \bibinfo
  {pages} {013428} (\bibinfo {year} {2013})}\BibitemShut {NoStop}%
\bibitem [{\citenamefont {Caycedo-Soler}\ \emph {et~al.}(2022)\citenamefont
  {Caycedo-Soler}, \citenamefont {Mattioni}, \citenamefont {Lim}, \citenamefont
  {Renger}, \citenamefont {Huelga},\ and\ \citenamefont
  {Plenio}}]{caycedosoler2021exact}%
  \BibitemOpen
  \bibfield  {author} {\bibinfo {author} {\bibfnamefont {F.}~\bibnamefont
  {Caycedo-Soler}}, \bibinfo {author} {\bibfnamefont {A.}~\bibnamefont
  {Mattioni}}, \bibinfo {author} {\bibfnamefont {J.}~\bibnamefont {Lim}},
  \bibinfo {author} {\bibfnamefont {T.}~\bibnamefont {Renger}}, \bibinfo
  {author} {\bibfnamefont {S.}~\bibnamefont {Huelga}},\ and\ \bibinfo {author}
  {\bibfnamefont {M.}~\bibnamefont {Plenio}},\ }\bibfield  {title} {\bibinfo
  {title} {Exact simulation of pigment-protein complexes unveils vibronic
  renormalization of electronic parameters in ultrafast spectroscopy},\
  }\href@noop {} {\bibfield  {journal} {\bibinfo  {journal} {Nature
  Communications}\ }\textbf {\bibinfo {volume} {13}},\ \bibinfo {pages} {1}
  (\bibinfo {year} {2022})}\BibitemShut {NoStop}%
\bibitem [{\citenamefont {Clark}\ \emph {et~al.}(2010)\citenamefont {Clark},
  \citenamefont {Prior}, \citenamefont {Hartmann}, \citenamefont {Jaksch},\
  and\ \citenamefont {Plenio}}]{plenioheisenberg}%
  \BibitemOpen
  \bibfield  {author} {\bibinfo {author} {\bibfnamefont {S.~R.}\ \bibnamefont
  {Clark}}, \bibinfo {author} {\bibfnamefont {J.}~\bibnamefont {Prior}},
  \bibinfo {author} {\bibfnamefont {M.~J.}\ \bibnamefont {Hartmann}}, \bibinfo
  {author} {\bibfnamefont {D.}~\bibnamefont {Jaksch}},\ and\ \bibinfo {author}
  {\bibfnamefont {M.~B.}\ \bibnamefont {Plenio}},\ }\bibfield  {title}
  {\bibinfo {title} {Exact matrix product solutions in the {Heisenberg} picture
  of an open quantum spin chain},\ }\href
  {http://stacks.iop.org/1367-2630/12/i=2/a=025005} {\bibfield  {journal}
  {\bibinfo  {journal} {New Journal of Physics}\ }\textbf {\bibinfo {volume}
  {12}},\ \bibinfo {pages} {025005} (\bibinfo {year} {2010})}\BibitemShut
  {NoStop}%
\bibitem [{\citenamefont {Werner}\ \emph {et~al.}(2016)\citenamefont {Werner},
  \citenamefont {Jaschke}, \citenamefont {Silvi}, \citenamefont {Kliesch},
  \citenamefont {Calarco}, \citenamefont {Eisert},\ and\ \citenamefont
  {Montangero}}]{PhysRevLett.116.237201}%
  \BibitemOpen
  \bibfield  {author} {\bibinfo {author} {\bibfnamefont {A.~H.}\ \bibnamefont
  {Werner}}, \bibinfo {author} {\bibfnamefont {D.}~\bibnamefont {Jaschke}},
  \bibinfo {author} {\bibfnamefont {P.}~\bibnamefont {Silvi}}, \bibinfo
  {author} {\bibfnamefont {M.}~\bibnamefont {Kliesch}}, \bibinfo {author}
  {\bibfnamefont {T.}~\bibnamefont {Calarco}}, \bibinfo {author} {\bibfnamefont
  {J.}~\bibnamefont {Eisert}},\ and\ \bibinfo {author} {\bibfnamefont
  {S.}~\bibnamefont {Montangero}},\ }\bibfield  {title} {\bibinfo {title}
  {Positive tensor network approach for simulating open quantum many-body
  systems},\ }\href {https://doi.org/10.1103/PhysRevLett.116.237201} {\bibfield
   {journal} {\bibinfo  {journal} {Phys. Rev. Lett.}\ }\textbf {\bibinfo
  {volume} {116}},\ \bibinfo {pages} {237201} (\bibinfo {year}
  {2016})}\BibitemShut {NoStop}%
\bibitem [{\citenamefont {Rosenbach}\ \emph {et~al.}(2016)\citenamefont
  {Rosenbach}, \citenamefont {Cerrillo}, \citenamefont {Huelga}, \citenamefont
  {Cao},\ and\ \citenamefont {Plenio}}]{Rosenbach_2016}%
  \BibitemOpen
  \bibfield  {author} {\bibinfo {author} {\bibfnamefont {R.}~\bibnamefont
  {Rosenbach}}, \bibinfo {author} {\bibfnamefont {J.}~\bibnamefont {Cerrillo}},
  \bibinfo {author} {\bibfnamefont {S.~F.}\ \bibnamefont {Huelga}}, \bibinfo
  {author} {\bibfnamefont {J.}~\bibnamefont {Cao}},\ and\ \bibinfo {author}
  {\bibfnamefont {M.~B.}\ \bibnamefont {Plenio}},\ }\bibfield  {title}
  {\bibinfo {title} {Efficient simulation of non-markovian system-environment
  interaction},\ }\href {https://doi.org/10.1088/1367-2630/18/2/023035}
  {\bibfield  {journal} {\bibinfo  {journal} {New Journal of Physics}\ }\textbf
  {\bibinfo {volume} {18}},\ \bibinfo {pages} {023035} (\bibinfo {year}
  {2016})}\BibitemShut {NoStop}%
\bibitem [{\citenamefont {Schr{\"o}der}\ \emph {et~al.}(2019)\citenamefont
  {Schr{\"o}der}, \citenamefont {Turban}, \citenamefont {Musser}, \citenamefont
  {Hine},\ and\ \citenamefont {Chin}}]{schroder2019tensor}%
  \BibitemOpen
  \bibfield  {author} {\bibinfo {author} {\bibfnamefont {F.~A.}\ \bibnamefont
  {Schr{\"o}der}}, \bibinfo {author} {\bibfnamefont {D.~H.}\ \bibnamefont
  {Turban}}, \bibinfo {author} {\bibfnamefont {A.~J.}\ \bibnamefont {Musser}},
  \bibinfo {author} {\bibfnamefont {N.~D.}\ \bibnamefont {Hine}},\ and\
  \bibinfo {author} {\bibfnamefont {A.~W.}\ \bibnamefont {Chin}},\ }\bibfield
  {title} {\bibinfo {title} {Tensor network simulation of multi-environmental
  open quantum dynamics via machine learning and entanglement
  renormalisation},\ }\href@noop {} {\bibfield  {journal} {\bibinfo  {journal}
  {Nature communications}\ }\textbf {\bibinfo {volume} {10}},\ \bibinfo {pages}
  {1} (\bibinfo {year} {2019})}\BibitemShut {NoStop}%
\bibitem [{\citenamefont {Somoza}\ \emph {et~al.}(2019)\citenamefont {Somoza},
  \citenamefont {Marty}, \citenamefont {Lim}, \citenamefont {Huelga},\ and\
  \citenamefont {Plenio}}]{PhysRevLett.123.100502}%
  \BibitemOpen
  \bibfield  {author} {\bibinfo {author} {\bibfnamefont {A.~D.}\ \bibnamefont
  {Somoza}}, \bibinfo {author} {\bibfnamefont {O.}~\bibnamefont {Marty}},
  \bibinfo {author} {\bibfnamefont {J.}~\bibnamefont {Lim}}, \bibinfo {author}
  {\bibfnamefont {S.~F.}\ \bibnamefont {Huelga}},\ and\ \bibinfo {author}
  {\bibfnamefont {M.~B.}\ \bibnamefont {Plenio}},\ }\bibfield  {title}
  {\bibinfo {title} {Dissipation-assisted matrix product factorization},\
  }\href {https://doi.org/10.1103/PhysRevLett.123.100502} {\bibfield  {journal}
  {\bibinfo  {journal} {Phys. Rev. Lett.}\ }\textbf {\bibinfo {volume} {123}},\
  \bibinfo {pages} {100502} (\bibinfo {year} {2019})}\BibitemShut {NoStop}%
\bibitem [{\citenamefont {Chernyak}\ and\ \citenamefont
  {Mukamel}(1996)}]{Chernyak:1996}%
  \BibitemOpen
  \bibfield  {author} {\bibinfo {author} {\bibfnamefont {V.}~\bibnamefont
  {Chernyak}}\ and\ \bibinfo {author} {\bibfnamefont {S.}~\bibnamefont
  {Mukamel}},\ }\bibfield  {title} {\bibinfo {title} {Collective coordinates
  for nuclear spectral densities in energy transfer and femtosecond
  spectroscopy of molecular aggregates},\ }\href
  {https://doi.org/10.1063/1.472302} {\bibfield  {journal} {\bibinfo  {journal}
  {J. Chem. Phys.}\ }\textbf {\bibinfo {volume} {105}},\ \bibinfo {pages}
  {4565} (\bibinfo {year} {1996})},\ \Eprint
  {https://arxiv.org/abs/https://doi.org/10.1063/1.472302}
  {https://doi.org/10.1063/1.472302} \BibitemShut {NoStop}%
\bibitem [{\citenamefont {Richter}\ and\ \citenamefont
  {Knorr}(2010)}]{RICHTER2010711}%
  \BibitemOpen
  \bibfield  {author} {\bibinfo {author} {\bibfnamefont {M.}~\bibnamefont
  {Richter}}\ and\ \bibinfo {author} {\bibfnamefont {A.}~\bibnamefont
  {Knorr}},\ }\bibfield  {title} {\bibinfo {title} {A time convolution less
  density matrix approach to the nonlinear optical response of a coupled
  system–bath complex},\ }\href
  {https://doi.org/https://doi.org/10.1016/j.aop.2009.12.008} {\bibfield
  {journal} {\bibinfo  {journal} {Annals of Physics}\ }\textbf {\bibinfo
  {volume} {325}},\ \bibinfo {pages} {711} (\bibinfo {year}
  {2010})}\BibitemShut {NoStop}%
\bibitem [{\citenamefont {Or{\'u}s}(2014)}]{orus2014practical}%
  \BibitemOpen
  \bibfield  {author} {\bibinfo {author} {\bibfnamefont {R.}~\bibnamefont
  {Or{\'u}s}},\ }\bibfield  {title} {\bibinfo {title} {A practical introduction
  to tensor networks: Matrix product states and projected entangled pair
  states},\ }\href@noop {} {\bibfield  {journal} {\bibinfo  {journal} {Annals
  of Physics}\ }\textbf {\bibinfo {volume} {349}},\ \bibinfo {pages} {117}
  (\bibinfo {year} {2014})}\BibitemShut {NoStop}%
\bibitem [{\citenamefont {Schollw{\"o}ck}(2011)}]{schollwock2011density}%
  \BibitemOpen
  \bibfield  {author} {\bibinfo {author} {\bibfnamefont {U.}~\bibnamefont
  {Schollw{\"o}ck}},\ }\bibfield  {title} {\bibinfo {title} {The density-matrix
  renormalization group in the age of matrix product states},\ }\href@noop {}
  {\bibfield  {journal} {\bibinfo  {journal} {Annals of physics}\ }\textbf
  {\bibinfo {volume} {326}},\ \bibinfo {pages} {96} (\bibinfo {year}
  {2011})}\BibitemShut {NoStop}%
\bibitem [{\citenamefont {Cirac}\ \emph {et~al.}(2017)\citenamefont {Cirac},
  \citenamefont {Pérez-García}, \citenamefont {Schuch},\ and\ \citenamefont
  {Verstraete}}]{CIRAC2017100}%
  \BibitemOpen
  \bibfield  {author} {\bibinfo {author} {\bibfnamefont {J.}~\bibnamefont
  {Cirac}}, \bibinfo {author} {\bibfnamefont {D.}~\bibnamefont
  {Pérez-García}}, \bibinfo {author} {\bibfnamefont {N.}~\bibnamefont
  {Schuch}},\ and\ \bibinfo {author} {\bibfnamefont {F.}~\bibnamefont
  {Verstraete}},\ }\bibfield  {title} {\bibinfo {title} {Matrix product density
  operators: Renormalization fixed points and boundary theories},\ }\href
  {https://doi.org/https://doi.org/10.1016/j.aop.2016.12.030} {\bibfield
  {journal} {\bibinfo  {journal} {Annals of Physics}\ }\textbf {\bibinfo
  {volume} {378}},\ \bibinfo {pages} {100 } (\bibinfo {year}
  {2017})}\BibitemShut {NoStop}%
\bibitem [{\citenamefont {Verstraete}\ and\ \citenamefont
  {Cirac}(2006)}]{verstraete2006matrix}%
  \BibitemOpen
  \bibfield  {author} {\bibinfo {author} {\bibfnamefont {F.}~\bibnamefont
  {Verstraete}}\ and\ \bibinfo {author} {\bibfnamefont {J.~I.}\ \bibnamefont
  {Cirac}},\ }\bibfield  {title} {\bibinfo {title} {Matrix product states
  represent ground states faithfully},\ }\href@noop {} {\bibfield  {journal}
  {\bibinfo  {journal} {Physical Review B}\ }\textbf {\bibinfo {volume} {73}},\
  \bibinfo {pages} {094423} (\bibinfo {year} {2006})}\BibitemShut {NoStop}%
\bibitem [{\citenamefont {Vidal}(2007)}]{vidal2007classical}%
  \BibitemOpen
  \bibfield  {author} {\bibinfo {author} {\bibfnamefont {G.}~\bibnamefont
  {Vidal}},\ }\bibfield  {title} {\bibinfo {title} {Classical simulation of
  infinite-size quantum lattice systems in one spatial dimension},\ }\href@noop
  {} {\bibfield  {journal} {\bibinfo  {journal} {Physical review letters}\
  }\textbf {\bibinfo {volume} {98}},\ \bibinfo {pages} {070201} (\bibinfo
  {year} {2007})}\BibitemShut {NoStop}%
\bibitem [{\citenamefont {Breuer}\ \emph {et~al.}(2002)\citenamefont {Breuer},
  \citenamefont {Petruccione} \emph {et~al.}}]{breuer2002theory}%
  \BibitemOpen
  \bibfield  {author} {\bibinfo {author} {\bibfnamefont {H.-P.}\ \bibnamefont
  {Breuer}}, \bibinfo {author} {\bibfnamefont {F.}~\bibnamefont {Petruccione}},
  \emph {et~al.},\ }\href@noop {} {\emph {\bibinfo {title} {The theory of open
  quantum systems}}}\ (\bibinfo  {publisher} {Oxford University Press on
  Demand},\ \bibinfo {year} {2002})\BibitemShut {NoStop}%
\bibitem [{\citenamefont {Garc{\'\i}a-Calder{\'o}n}\ and\ \citenamefont
  {Peierls}(1976)}]{garcia1976resonant}%
  \BibitemOpen
  \bibfield  {author} {\bibinfo {author} {\bibfnamefont {G.}~\bibnamefont
  {Garc{\'\i}a-Calder{\'o}n}}\ and\ \bibinfo {author} {\bibfnamefont
  {R.}~\bibnamefont {Peierls}},\ }\bibfield  {title} {\bibinfo {title}
  {Resonant states and their uses},\ }\href@noop {} {\bibfield  {journal}
  {\bibinfo  {journal} {Nuclear Physics A}\ }\textbf {\bibinfo {volume}
  {265}},\ \bibinfo {pages} {443} (\bibinfo {year} {1976})}\BibitemShut
  {NoStop}%
\bibitem [{\citenamefont {Lee}\ \emph {et~al.}(1999)\citenamefont {Lee},
  \citenamefont {Leung},\ and\ \citenamefont {Pang}}]{lee1999dyadic}%
  \BibitemOpen
  \bibfield  {author} {\bibinfo {author} {\bibfnamefont {K.}~\bibnamefont
  {Lee}}, \bibinfo {author} {\bibfnamefont {P.}~\bibnamefont {Leung}},\ and\
  \bibinfo {author} {\bibfnamefont {K.}~\bibnamefont {Pang}},\ }\bibfield
  {title} {\bibinfo {title} {Dyadic formulation of morphology-dependent
  resonances. i. completeness relation},\ }\href@noop {} {\bibfield  {journal}
  {\bibinfo  {journal} {JOSA B}\ }\textbf {\bibinfo {volume} {16}},\ \bibinfo
  {pages} {1409} (\bibinfo {year} {1999})}\BibitemShut {NoStop}%
\bibitem [{\citenamefont {Muljarov}\ \emph {et~al.}(2011)\citenamefont
  {Muljarov}, \citenamefont {Langbein},\ and\ \citenamefont
  {Zimmermann}}]{muljarov2011brillouin}%
  \BibitemOpen
  \bibfield  {author} {\bibinfo {author} {\bibfnamefont {E.~A.}\ \bibnamefont
  {Muljarov}}, \bibinfo {author} {\bibfnamefont {W.}~\bibnamefont {Langbein}},\
  and\ \bibinfo {author} {\bibfnamefont {R.}~\bibnamefont {Zimmermann}},\
  }\bibfield  {title} {\bibinfo {title} {Brillouin-wigner perturbation theory
  in open electromagnetic systems},\ }\href@noop {} {\bibfield  {journal}
  {\bibinfo  {journal} {EPL (Europhysics Letters)}\ }\textbf {\bibinfo {volume}
  {92}},\ \bibinfo {pages} {50010} (\bibinfo {year} {2011})}\BibitemShut
  {NoStop}%
\bibitem [{\citenamefont {Kristensen}\ \emph {et~al.}(2012)\citenamefont
  {Kristensen}, \citenamefont {Van~Vlack},\ and\ \citenamefont
  {Hughes}}]{kristensen2012generalized}%
  \BibitemOpen
  \bibfield  {author} {\bibinfo {author} {\bibfnamefont {P.~T.}\ \bibnamefont
  {Kristensen}}, \bibinfo {author} {\bibfnamefont {C.}~\bibnamefont
  {Van~Vlack}},\ and\ \bibinfo {author} {\bibfnamefont {S.}~\bibnamefont
  {Hughes}},\ }\bibfield  {title} {\bibinfo {title} {Generalized effective mode
  volume for leaky optical cavities},\ }\href@noop {} {\bibfield  {journal}
  {\bibinfo  {journal} {Optics letters}\ }\textbf {\bibinfo {volume} {37}},\
  \bibinfo {pages} {1649} (\bibinfo {year} {2012})}\BibitemShut {NoStop}%
\bibitem [{\citenamefont {Sauvan}\ \emph {et~al.}(2013)\citenamefont {Sauvan},
  \citenamefont {Hugonin}, \citenamefont {Maksymov},\ and\ \citenamefont
  {Lalanne}}]{sauvan2013theory}%
  \BibitemOpen
  \bibfield  {author} {\bibinfo {author} {\bibfnamefont {C.}~\bibnamefont
  {Sauvan}}, \bibinfo {author} {\bibfnamefont {J.-P.}\ \bibnamefont {Hugonin}},
  \bibinfo {author} {\bibfnamefont {I.~S.}\ \bibnamefont {Maksymov}},\ and\
  \bibinfo {author} {\bibfnamefont {P.}~\bibnamefont {Lalanne}},\ }\bibfield
  {title} {\bibinfo {title} {Theory of the spontaneous optical emission of
  nanosize photonic and plasmon resonators},\ }\href@noop {} {\bibfield
  {journal} {\bibinfo  {journal} {Physical Review Letters}\ }\textbf {\bibinfo
  {volume} {110}},\ \bibinfo {pages} {237401} (\bibinfo {year}
  {2013})}\BibitemShut {NoStop}%
\bibitem [{\citenamefont {Franke}\ \emph {et~al.}(2019)\citenamefont {Franke},
  \citenamefont {Hughes}, \citenamefont {Dezfouli}, \citenamefont {Kristensen},
  \citenamefont {Busch}, \citenamefont {Knorr},\ and\ \citenamefont
  {Richter}}]{franke2019quantization}%
  \BibitemOpen
  \bibfield  {author} {\bibinfo {author} {\bibfnamefont {S.}~\bibnamefont
  {Franke}}, \bibinfo {author} {\bibfnamefont {S.}~\bibnamefont {Hughes}},
  \bibinfo {author} {\bibfnamefont {M.~K.}\ \bibnamefont {Dezfouli}}, \bibinfo
  {author} {\bibfnamefont {P.~T.}\ \bibnamefont {Kristensen}}, \bibinfo
  {author} {\bibfnamefont {K.}~\bibnamefont {Busch}}, \bibinfo {author}
  {\bibfnamefont {A.}~\bibnamefont {Knorr}},\ and\ \bibinfo {author}
  {\bibfnamefont {M.}~\bibnamefont {Richter}},\ }\bibfield  {title} {\bibinfo
  {title} {Quantization of quasinormal modes for open cavities and plasmonic
  cavity quantum electrodynamics},\ }\href@noop {} {\bibfield  {journal}
  {\bibinfo  {journal} {Physical review letters}\ }\textbf {\bibinfo {volume}
  {122}},\ \bibinfo {pages} {213901} (\bibinfo {year} {2019})}\BibitemShut
  {NoStop}%
\bibitem [{\citenamefont {Kristensen}\ \emph {et~al.}(2020)\citenamefont
  {Kristensen}, \citenamefont {Herrmann}, \citenamefont {Intravaia},\ and\
  \citenamefont {Busch}}]{kristensen2020modeling}%
  \BibitemOpen
  \bibfield  {author} {\bibinfo {author} {\bibfnamefont {P.~T.}\ \bibnamefont
  {Kristensen}}, \bibinfo {author} {\bibfnamefont {K.}~\bibnamefont
  {Herrmann}}, \bibinfo {author} {\bibfnamefont {F.}~\bibnamefont
  {Intravaia}},\ and\ \bibinfo {author} {\bibfnamefont {K.}~\bibnamefont
  {Busch}},\ }\bibfield  {title} {\bibinfo {title} {Modeling electromagnetic
  resonators using quasinormal modes},\ }\href@noop {} {\bibfield  {journal}
  {\bibinfo  {journal} {Advances in Optics and Photonics}\ }\textbf {\bibinfo
  {volume} {12}},\ \bibinfo {pages} {612} (\bibinfo {year} {2020})}\BibitemShut
  {NoStop}%
\bibitem [{\citenamefont {Bello}\ \emph {et~al.}(2019)\citenamefont {Bello},
  \citenamefont {Platero}, \citenamefont {Cirac},\ and\ \citenamefont
  {Gonz{\'a}lez-Tudela}}]{bello2019unconventional}%
  \BibitemOpen
  \bibfield  {author} {\bibinfo {author} {\bibfnamefont {M.}~\bibnamefont
  {Bello}}, \bibinfo {author} {\bibfnamefont {G.}~\bibnamefont {Platero}},
  \bibinfo {author} {\bibfnamefont {J.~I.}\ \bibnamefont {Cirac}},\ and\
  \bibinfo {author} {\bibfnamefont {A.}~\bibnamefont {Gonz{\'a}lez-Tudela}},\
  }\bibfield  {title} {\bibinfo {title} {Unconventional quantum optics in
  topological waveguide qed},\ }\href@noop {} {\bibfield  {journal} {\bibinfo
  {journal} {Science advances}\ }\textbf {\bibinfo {volume} {5}},\ \bibinfo
  {pages} {eaaw0297} (\bibinfo {year} {2019})}\BibitemShut {NoStop}%
\bibitem [{\citenamefont {Hughes}\ and\ \citenamefont
  {Agarwal}(2017)}]{hughes2017anisotropy}%
  \BibitemOpen
  \bibfield  {author} {\bibinfo {author} {\bibfnamefont {S.}~\bibnamefont
  {Hughes}}\ and\ \bibinfo {author} {\bibfnamefont {G.~S.}\ \bibnamefont
  {Agarwal}},\ }\bibfield  {title} {\bibinfo {title} {Anisotropy-induced
  quantum interference and population trapping between orthogonal quantum dot
  exciton states in semiconductor cavity systems},\ }\href@noop {} {\bibfield
  {journal} {\bibinfo  {journal} {Physical review letters}\ }\textbf {\bibinfo
  {volume} {118}},\ \bibinfo {pages} {063601} (\bibinfo {year}
  {2017})}\BibitemShut {NoStop}%
\bibitem [{\citenamefont {Grimsmo}(2015)}]{grimsmo2015time}%
  \BibitemOpen
  \bibfield  {author} {\bibinfo {author} {\bibfnamefont {A.~L.}\ \bibnamefont
  {Grimsmo}},\ }\bibfield  {title} {\bibinfo {title} {Time-delayed quantum
  feedback control},\ }\href@noop {} {\bibfield  {journal} {\bibinfo  {journal}
  {Physical review letters}\ }\textbf {\bibinfo {volume} {115}},\ \bibinfo
  {pages} {060402} (\bibinfo {year} {2015})}\BibitemShut {NoStop}%
\bibitem [{\citenamefont {N{\'e}met}\ \emph {et~al.}(2019)\citenamefont
  {N{\'e}met}, \citenamefont {Carmele}, \citenamefont {Parkins},\ and\
  \citenamefont {Knorr}}]{nemet2019comparison}%
  \BibitemOpen
  \bibfield  {author} {\bibinfo {author} {\bibfnamefont {N.}~\bibnamefont
  {N{\'e}met}}, \bibinfo {author} {\bibfnamefont {A.}~\bibnamefont {Carmele}},
  \bibinfo {author} {\bibfnamefont {S.}~\bibnamefont {Parkins}},\ and\ \bibinfo
  {author} {\bibfnamefont {A.}~\bibnamefont {Knorr}},\ }\bibfield  {title}
  {\bibinfo {title} {Comparison between continuous-and discrete-mode coherent
  feedback for the jaynes-cummings model},\ }\href@noop {} {\bibfield
  {journal} {\bibinfo  {journal} {Physical Review A}\ }\textbf {\bibinfo
  {volume} {100}},\ \bibinfo {pages} {023805} (\bibinfo {year}
  {2019})}\BibitemShut {NoStop}%
\bibitem [{\citenamefont {Barkemeyer}\ \emph {et~al.}(2020)\citenamefont
  {Barkemeyer}, \citenamefont {Finsterh{\"o}lzl}, \citenamefont {Knorr},\ and\
  \citenamefont {Carmele}}]{barkemeyer2020revisiting}%
  \BibitemOpen
  \bibfield  {author} {\bibinfo {author} {\bibfnamefont {K.}~\bibnamefont
  {Barkemeyer}}, \bibinfo {author} {\bibfnamefont {R.}~\bibnamefont
  {Finsterh{\"o}lzl}}, \bibinfo {author} {\bibfnamefont {A.}~\bibnamefont
  {Knorr}},\ and\ \bibinfo {author} {\bibfnamefont {A.}~\bibnamefont
  {Carmele}},\ }\bibfield  {title} {\bibinfo {title} {Revisiting quantum
  feedback control: disentangling the feedback-induced phase from the
  corresponding amplitude},\ }\href@noop {} {\bibfield  {journal} {\bibinfo
  {journal} {Advanced Quantum Technologies}\ }\textbf {\bibinfo {volume} {3}},\
  \bibinfo {pages} {1900078} (\bibinfo {year} {2020})}\BibitemShut {NoStop}%
\bibitem [{\citenamefont {Finsterh{\"o}lzl}\ \emph {et~al.}(2020)\citenamefont
  {Finsterh{\"o}lzl}, \citenamefont {Katzer},\ and\ \citenamefont
  {Carmele}}]{finsterholzl2020nonequilibrium}%
  \BibitemOpen
  \bibfield  {author} {\bibinfo {author} {\bibfnamefont {R.}~\bibnamefont
  {Finsterh{\"o}lzl}}, \bibinfo {author} {\bibfnamefont {M.}~\bibnamefont
  {Katzer}},\ and\ \bibinfo {author} {\bibfnamefont {A.}~\bibnamefont
  {Carmele}},\ }\bibfield  {title} {\bibinfo {title} {Nonequilibrium
  non-markovian steady states in open quantum many-body systems: Persistent
  oscillations in heisenberg quantum spin chains},\ }\href@noop {} {\bibfield
  {journal} {\bibinfo  {journal} {Physical Review B}\ }\textbf {\bibinfo
  {volume} {102}},\ \bibinfo {pages} {174309} (\bibinfo {year}
  {2020})}\BibitemShut {NoStop}%
\bibitem [{\citenamefont {Lalanne}\ \emph {et~al.}(2018)\citenamefont
  {Lalanne}, \citenamefont {Yan}, \citenamefont {Vynck}, \citenamefont
  {Sauvan},\ and\ \citenamefont {Hugonin}}]{lalanne2018light}%
  \BibitemOpen
  \bibfield  {author} {\bibinfo {author} {\bibfnamefont {P.}~\bibnamefont
  {Lalanne}}, \bibinfo {author} {\bibfnamefont {W.}~\bibnamefont {Yan}},
  \bibinfo {author} {\bibfnamefont {K.}~\bibnamefont {Vynck}}, \bibinfo
  {author} {\bibfnamefont {C.}~\bibnamefont {Sauvan}},\ and\ \bibinfo {author}
  {\bibfnamefont {J.-P.}\ \bibnamefont {Hugonin}},\ }\bibfield  {title}
  {\bibinfo {title} {Light interaction with photonic and plasmonic
  resonances},\ }\href@noop {} {\bibfield  {journal} {\bibinfo  {journal}
  {Laser \& Photonics Reviews}\ }\textbf {\bibinfo {volume} {12}},\ \bibinfo
  {pages} {1700113} (\bibinfo {year} {2018})}\BibitemShut {NoStop}%
\bibitem [{\citenamefont {Ge}\ \emph {et~al.}(2014)\citenamefont {Ge},
  \citenamefont {Kristensen}, \citenamefont {Young},\ and\ \citenamefont
  {Hughes}}]{ge2014quasinormal}%
  \BibitemOpen
  \bibfield  {author} {\bibinfo {author} {\bibfnamefont {R.-C.}\ \bibnamefont
  {Ge}}, \bibinfo {author} {\bibfnamefont {P.~T.}\ \bibnamefont {Kristensen}},
  \bibinfo {author} {\bibfnamefont {J.~F.}\ \bibnamefont {Young}},\ and\
  \bibinfo {author} {\bibfnamefont {S.}~\bibnamefont {Hughes}},\ }\bibfield
  {title} {\bibinfo {title} {Quasinormal mode approach to modelling
  light-emission and propagation in nanoplasmonics},\ }\href@noop {} {\bibfield
   {journal} {\bibinfo  {journal} {New Journal of Physics}\ }\textbf {\bibinfo
  {volume} {16}},\ \bibinfo {pages} {113048} (\bibinfo {year}
  {2014})}\BibitemShut {NoStop}%
\bibitem [{\citenamefont {Franke}\ \emph
  {et~al.}(2020{\natexlab{a}})\citenamefont {Franke}, \citenamefont {Ren},
  \citenamefont {Hughes},\ and\ \citenamefont
  {Richter}}]{franke2020fluctuation}%
  \BibitemOpen
  \bibfield  {author} {\bibinfo {author} {\bibfnamefont {S.}~\bibnamefont
  {Franke}}, \bibinfo {author} {\bibfnamefont {J.}~\bibnamefont {Ren}},
  \bibinfo {author} {\bibfnamefont {S.}~\bibnamefont {Hughes}},\ and\ \bibinfo
  {author} {\bibfnamefont {M.}~\bibnamefont {Richter}},\ }\bibfield  {title}
  {\bibinfo {title} {Fluctuation-dissipation theorem and fundamental photon
  commutation relations in lossy nanostructures using quasinormal modes},\
  }\href@noop {} {\bibfield  {journal} {\bibinfo  {journal} {Physical Review
  Research}\ }\textbf {\bibinfo {volume} {2}},\ \bibinfo {pages} {033332}
  (\bibinfo {year} {2020}{\natexlab{a}})}\BibitemShut {NoStop}%
\bibitem [{\citenamefont {Gruner}\ and\ \citenamefont
  {Welsch}(1996)}]{PhysRevA.53.1818}%
  \BibitemOpen
  \bibfield  {author} {\bibinfo {author} {\bibfnamefont {T.}~\bibnamefont
  {Gruner}}\ and\ \bibinfo {author} {\bibfnamefont {D.-G.}\ \bibnamefont
  {Welsch}},\ }\bibfield  {title} {\bibinfo {title} {Green-function approach to
  the radiation-field quantization for homogeneous and inhomogeneous
  kramers-kronig dielectrics},\ }\href
  {https://doi.org/10.1103/PhysRevA.53.1818} {\bibfield  {journal} {\bibinfo
  {journal} {Phys. Rev. A}\ }\textbf {\bibinfo {volume} {53}},\ \bibinfo
  {pages} {1818} (\bibinfo {year} {1996})}\BibitemShut {NoStop}%
\bibitem [{\citenamefont {Franke}\ \emph
  {et~al.}(2020{\natexlab{b}})\citenamefont {Franke}, \citenamefont {Richter},
  \citenamefont {Ren}, \citenamefont {Knorr},\ and\ \citenamefont
  {Hughes}}]{franke2020quantized}%
  \BibitemOpen
  \bibfield  {author} {\bibinfo {author} {\bibfnamefont {S.}~\bibnamefont
  {Franke}}, \bibinfo {author} {\bibfnamefont {M.}~\bibnamefont {Richter}},
  \bibinfo {author} {\bibfnamefont {J.}~\bibnamefont {Ren}}, \bibinfo {author}
  {\bibfnamefont {A.}~\bibnamefont {Knorr}},\ and\ \bibinfo {author}
  {\bibfnamefont {S.}~\bibnamefont {Hughes}},\ }\bibfield  {title} {\bibinfo
  {title} {Quantized quasinormal-mode description of nonlinear cavity-qed
  effects from coupled resonators with a fano-like resonance},\ }\href@noop {}
  {\bibfield  {journal} {\bibinfo  {journal} {Physical Review Research}\
  }\textbf {\bibinfo {volume} {2}},\ \bibinfo {pages} {033456} (\bibinfo {year}
  {2020}{\natexlab{b}})}\BibitemShut {NoStop}%
\bibitem [{\citenamefont {Mukamel}(1999)}]{mukamel1999principles}%
  \BibitemOpen
  \bibfield  {author} {\bibinfo {author} {\bibfnamefont {S.}~\bibnamefont
  {Mukamel}},\ }\href@noop {} {\emph {\bibinfo {title} {Principles of nonlinear
  optical spectroscopy}}},\ \bibinfo {number} {6}\ (\bibinfo  {publisher}
  {Oxford University Press on Demand},\ \bibinfo {year} {1999})\BibitemShut
  {NoStop}%
\end{thebibliography}%

\end{document}